\documentclass[%
 reprint,
 superscriptaddress,
 amsmath,amssymb,
 aps,
 floatfix,
]{revtex4-1}

\usepackage{graphicx}
\usepackage{dcolumn}
\usepackage{bm}
\usepackage{physics}
\usepackage{mathtools}
\usepackage{amsfonts}
\usepackage{amssymb}
\usepackage{amsmath}
\usepackage[caption=false,labelformat=empty]{subfig}
\usepackage{diagbox}
\usepackage{hyperref}
\hypersetup{linktocpage,colorlinks,citecolor={blue},pdfdisplaydoctitle=true,pdfpagemode=UseOutlines,bookmarksnumbered=true}

\begin{document}


\title{Real-time dynamics in 2+1d compact QED using complex periodic Gaussian states }

\author{Julian Bender}
\author{Patrick Emonts}
\affiliation{Max-Planck Institute of Quantum Optics, Hans-Kopfermann-Str. 1, 85748 Garching, Germany}
\affiliation{Munich Center for Quantum Science and Technology (MCQST), Schellingstr. 4, D-80799 München}
\author{Erez Zohar}
\affiliation{Racah Institute of Physics, The Hebrew University of Jerusalem, Givat Ram, Jerusalem 91904, Israel}
\author{J. Ignacio Cirac}
\affiliation{Max-Planck Institute of Quantum Optics, Hans-Kopfermann-Str. 1, 85748 Garching, Germany}
\affiliation{Munich Center for Quantum Science and Technology (MCQST), Schellingstr. 4, D-80799 München}

\date{\today}

\begin{abstract}
We introduce a class of variational states to study ground state properties and real-time dynamics in (2+1)-dimensional compact QED. 
These are based on complex Gaussian states which are made periodic in order to account for the compact nature of the $U(1)$ gauge field. 
Since the evaluation of expectation values involves infinite sums, we present an approximation scheme for the whole variational manifold. 
We calculate the ground state energy density for lattice sizes up to $20 \times 20$ and extrapolate to the thermodynamic limit for the whole coupling region. 
Additionally, we study the string tension both by fitting the potential between two static charges and by fitting the exponential decay of spatial Wilson loops. 
As the ansatz does not require a truncation in the local Hilbert spaces, we analyze truncation effects which are present in other approaches. 
The variational states are benchmarked against exact solutions known for the one plaquette case and exact diagonalization results for a $\mathbb{Z}_3$ lattice gauge theory. 
Using the time-dependent variational principle, we study real-time dynamics after various global quenches, e.g. the time evolution of a strongly confined electric field between two charges after a quench to the weak-coupling regime. 
Up to the points where finite size effects start to play a role, we observe equilibrating behavior.
\end{abstract}

\maketitle


\section{\label{sec:intro}Introduction}

Gauge theories are of paramount importance in fundamental physics.
Its most prominent example, the standard model of particle physics, describes electromagnetic, weak and strong interactions. 
In some regimes, interactions can be treated in terms of perturbative expansions. 
However, since the coupling in quantum field theories is typically scale-dependent, there are regimes (e.g. low-energy QCD) where non-perturbative methods are required~\cite{peskin1995introduction,gross1973asymptotically}. 

Lattice gauge theory is a gauge invariant lattice regularization of gauge theories, in which either spacetime~\cite{wilson1974confinement} or space~\cite{kogut1975hamiltonian} is discretized.
This has allowed to uncover many interesting features of non-perturbative quantum field theories, in particular using Monte-Carlo simulations~\cite{aoki2014review}. 
Nevertheless, certain aspects are difficult to study within this framework, e.g. fermionic theories with finite chemical potentials may suffer from the sign problem~\cite{troyer2005computational} and time dynamics are difficult to access as Monte-Carlo simulations require a formulation in Euclidean spacetime. 

One class of approaches to these problems is based on a Hamiltonian formulation of lattice gauge theories, first proposed by Kogut and Susskind~\cite{kogut1975hamiltonian}. 
Other formulations in the Hamiltonian picture include the quantum link model~\cite{horn1981finite,orland1990lattice,chandrasekharan1997quantum,brower1999qcd} or the prepotential approach~\cite{mathur2005harmonic}. 
It has been shown that these Hamiltonians or truncations~\cite{zohar2015formulation} thereof can be mapped to Hamiltonians of quantum devices (e.g. ultracold atoms, trapped ions or superconducting qubits) in order to study such theories by quantum simulation~\cite{wiese2014towards,zohar2016quantum,dalmonte2016lattice, byrnes2006simulating}. 
Another option is to study the Hamiltonian by designing appropriate variational ansatz states which are both efficiently tractable and capture the most relevant features of the theory. 

Both ideas have been successfully applied to one-dimensional theories. The implementation of  quantum simulators has been demonstrated using trapped ions~\cite{martinez2016real} and ultracold atoms~\cite{gorg2019realization,schweizer2019floquet,yang2020observation,mil2020scalable}. 
On the numerical side, there has been a lot of success in applying matrix product state (MPS) methods to (1+1)-dimensional Abelian and non-Abelian lattice gauge theories~\cite{banuls2017density,buyens2014matrix,buyens2016hamiltonian,buyens2017real,kuhn2015non,banuls2013mass,banuls2015thermal,banuls2017efficient,pichler2016real,silvi2017finite,silvi2019tensor,silvi2014lattice,bruckmann2019nonlinear,funcke2020topological,rico2014tensor}, enabling the study of finite chemical potential scenarios and out-of-equilibrium dynamics which would not have been accessible in Monte-Carlo simulations of Euclidean lattice gauge theory. 
Also some generalizations of Gaussian states have proven to be suitable for these theories~\cite{sala2018variational}. 

The situation becomes more challenging in higher spatial dimensions, in particular due to appearance of magnetic interactions, leading to four-body plaquette terms on the lattice. 
There have been ideas on how to overcome this problem in quantum simulators (either by employing a digital~\cite{tagliacozzo2013simulation,tagliacozzo2013optical,zohar2017digital,zohar2017digital,bender2018digital} or an analog simulation scheme~\cite{zohar2013quantum}) but so far they are out of experimental reach. 
On the numerical side, tensor network methods in 2+1d have been applied to pure gauge theories~\cite{tagliacozzo2014tensor} and for studying U(1) ground states in quantum link models~\cite{tschirsich2019phase,felser2019two}. It has also been shown that fermionic Gaussian projected entangled pair states can be gauged~\cite{zohar2015fermionic} and serve as numerical ansatz states for lattice gauge theories, admitting a sign-problem free Monte-Carlo contraction scheme~\cite{zohar2018combining}. 

In this work, we study (2+1)-dimensional compact quantum electrodynamics (compact QED). 
It is a good starting point for the study of higher dimensional lattice gauge theories since it shares some features with (3+1)-dimensional Quantum Chromodynamics (QCD), e.g. that it is in a confined phase for all values of the coupling constant~\cite{polyakov1977quark}. 
To access physics which is difficult to simulate with Monte-Carlo simulation of Euclidean lattice gauge theories, we not only study ground state properties but also non-equilibrium physics, namely real-time dynamics after a quantum quench. 

Since exact diagonalization (ED) methods become infeasible in higher dimensions for reasonable system sizes, in particular due to the infinite local Hilbert space of the gauge field, it seems unavoidable to use variational techniques (in 1+1d the infinite dimension can be avoided either by integrating out the gauge field nonlocally \cite{hamer1997series,bringoltz2009volume,banuls2017efficient} or by using the natural restriction of gauge symmetry which makes the dimensions finite \cite{kasper2020jaynes-cummings}). 

We choose to work with complex periodic Gaussian states, a generalization of periodic Gaussian states, first proposed in~\cite{drell1979quantum} to prove confinement in the weak-coupling limit of 2+1d compact QED, thus establishing the existence of one confining phase for all couplings also in the Hamiltonian picture (after it had been proven in the action formalism~\cite{polyakov1977quark}).
As expectation values with respect to periodic Gaussian states cannot be evaluated analytically, the authors of reference~\cite{drell1979quantum} used Feynman diagram techniques to evaluate all relevant quantities in the weak-coupling regime.
In contrast to that approach, we develop a numerical approximation scheme to evaluate these states for the whole coupling region.
By extending the variational manifold to complex periodic Gaussian states we are also able to account for real-time dynamics. 
One appealing feature of these states is that they do not require any truncation in Hilbert space which allows us to study truncation effects which are required in other approaches and give estimates in which coupling regimes they are justified. 

The manuscript is structured as follows: In Sec.~\ref{model}, we introduce the model and the variational ansatz including a scheme for its numerical evaluation.
In the first part of Sec.~\ref{static}, we study ground state energy density and string tension over the whole coupling region.
In the second part, we investigate truncation effects by comparing the variational ground state energy with exact diagonalization results where the local Hilbert space is truncated in the electric basis. 
In Sec.~\ref{dynamic}, we study real-time dynamics after a quantum quench using the time-dependent variational principle.
In Sec.~\ref{conclusion}, we conclude.

\section{\label{model} Model and variational ansatz} 
                       
\subsection{(2+1)-dimensional compact QED} \label{modelsection}
We define the theory of (2+1)-dimensional compact QED on a square lattice of extent $L \times L$ with periodic boundary conditions. 
The gauge fields reside on the links;  $U_{\mathbf{x},i}$ denotes the gauge field operator on the link emanating from site $\mathbf{x}$ in direction $\mathbf{e}_{i}$. 
The Hamiltonian in lattice units takes the following form, originally proposed by Kogut and Susskind~\cite{kogut1975hamiltonian}: 
\begin{equation} \label{kogutsusskind}
H_{KS}= \frac{g^2}{2}\sum_{\mathbf{x},i} E^2_{\mathbf{x},i} + \frac{1}{2 g^2}  \sum_{\mathbf{p}} 2 - ( U_{\mathbf{p}} +  U_{\mathbf{p}}^{\dagger})
\end{equation} 
with $g^2$ being the coupling constant and $U_{\mathbf{p}} \equiv U_{\mathbf{x},1} U_{\mathbf{x}+\mathbf{e_1},2} U_{\mathbf{x}+\mathbf{e_2},1}^{\dagger} U_{\mathbf{x},2}^{\dagger}$ where $\mathbf{x}$ is the bottom left corner of plaquette $\mathbf{p}$. 
$U_{\mathbf{x},i}$ is in the fundamental representation of $U(1)$, it can also be written in terms of an angle $\theta_{\mathbf{x},i}$, $U_{\mathbf{x},i}=e^{i \theta_{\mathbf{x},i}}$ with $ -\pi < \theta_{\mathbf{x},i} \leq \pi$. 
The restriction of the gauge field to this compact interval is the reason why the model is called compact QED and why it exhibits interesting features such as confinement in contrast to the non-compact theory~\cite{ben1979confinement}. 
$E_{\mathbf{x},i}$ is the electric field operator fulfilling the following commutation relations: 
\begin{equation}
    \begin{aligned}
    	[E_{\mathbf{x},i},U_{\mathbf{y},j}]&= \delta_{\mathbf{x},\mathbf{y}} \delta_{i,j} U_{\mathbf{x},i} \\
    	[\theta_{\mathbf{x},i},E_{\mathbf{y},j}]&= i \delta_{\mathbf{x},\mathbf{y}} \delta_{i,j} 
    \end{aligned}
\end{equation}
Since we work in the temporal gauge, there is a residual spatial gauge symmetry defined by the Gauss law operators $G_{\mathbf{x}}$. 
All physical states need to be eigenstates of them: 
\begin{align}
G_{\mathbf{x}} \ket{\mathrm{phys}}=\sum_{i=1}^{2} \left( E_{\mathbf{x},i} - E_{\mathbf{x}-\mathbf{e_{i}},i}\right) \ket{\mathrm{phys}} = Q_{\mathbf{x}}\ket{\mathrm{phys}} \hspace{5pt} \forall \hspace{1pt} \mathbf{x}
\end{align}
where the eigenvalue $Q_{\mathbf{x}}$  gives the static charge configuration at $\mathbf{x}$. 

These local constraints put quite severe restrictions on the choice of variational states. 
Following~\cite{drell1979quantum}, we thus change to variables where gauge invariance is already incorporated (at least up to a global constraint). 
This can be achieved by splitting the electric field $E_{\mathbf{x},i}$ into its transversal part $E_{\mathbf{x},i}^T$, which is dynamical, and a longitudinal part $E_{\mathbf{x},i}^L$ which is fixed by the static charge configuration. 
Since the transversal part of the electric field can be expressed by a plaquette field $L_{\mathbf{p}}$ (the lattice analogue of a solenoidal vector field), the remaining dynamical degrees of freedom $\{ L_{\mathbf{p}},U_{\mathbf{p}}=e^{i \theta_{\mathbf{p}}} \}$ reside on plaquettes, having the same Hilbert space structure and fulfilling the same commutation relations as the link variables: 

\begin{equation}
\begin{aligned}
[L_{\mathbf{p}},U_{\mathbf{p'}}]&= \delta_{\mathbf{p},\mathbf{p'}}  U_{\mathbf{p'}} \\
[\theta_{\mathbf{p}},L_{\mathbf{p'}}]&=i \delta_{\mathbf{p},\mathbf{p'}}  
\end{aligned}
\end{equation}
The operator $U_{\mathbf{p}}$ creates an electric flux excitation around plaquette $\mathbf{p}$. However, to construct all possible gauge-invariant flux configurations two global non-contractible flux loops around the torus (one for each spatial direction) are required, their operators are denoted as $ \{ \theta_{1}, L_{1} \}$ and $\{ \theta_{2}, L_{2} \}$ specifying the topological sector of the flux configuration. 
$L_1$ and $L_2$ commute with the Hamiltonian and we will restrict ourselves to the topological sector with $L_{1}=L_{2}=0$ which corresponds to no electric flux loops winding around the torus. 
For more details see~\cite{kaplan2018gauss} or Appendix~\ref{appformulation}.
Writing the Hamiltonian in terms of these new variables, reads 
\begin{equation}
    \label{Hamiltonianref}
    \begin{aligned}
    H_{KS}=&E_C+\frac{1}{g^2}  \sum_{\mathbf{p}} (1- \cos{\theta_{\mathbf{p}}}) \\
    +&  \frac{g^2}{2}\sum_{\mathbf{p}} \sum_{i=1}^{2} \left(L_\mathbf{p}-L_{\mathbf{p}-\mathbf{e}_{i}}+\epsilon_{\mathbf{p}}-\epsilon_{\mathbf{p}-\mathbf{e}_{i}}\right)^2
    \end{aligned}
\end{equation}
where $E_C$ is an energy offset given by the lattice Coulomb energy and $\epsilon_{\mathbf{p}}$ accounts for the transversal part of the electric field caused by the static charges only, i.e $\epsilon_{\mathbf{p}}=0$ in case of no static charges.
Even in this formulation there is one remaining global constraint left which is intuitively clear since raising the electric flux around all plaquettes should return the same state due to the periodic boundary conditions. Thus, 

\begin{align} \label{globalconstraint}
\prod_{\mathbf{p}} U_{\mathbf{p}} \ket{\mathrm{phys}}=\ket{\mathrm{phys}} 
\end{align}
For details on this formulation, we refer the reader to~\cite{drell1979quantum,kaplan2018gauss}.  A rigorous derivation of eq.~\eqref{Hamiltonianref} from eq.~\eqref{kogutsusskind} and an explicit formula for the calculation of $\epsilon_{\mathbf{p}}$ and $E_{C}$ can be found in Appendix~\ref{appformulation}.

\subsection{The variational ansatz} \label{variationalansatz}

We formulate our variational ansatz states in terms of the $\theta_{\mathbf{p}}$-variables defined above such that it only needs to fulfill the global constraint~\eqref{globalconstraint}.
Starting from periodic Gaussian states introduced in~\cite{drell1979quantum}, we extend the variational wavefunction to have an imaginary part in order to account for real-time dynamics. The ansatz is based on a complex Gaussian state:
\begin{equation}
\Psi_{CG}(\{ x_{\mathbf{p}} \})\equiv e^{-\frac{1}{2} \sum_{\mathbf{p},\mathbf{p'}} x_{\mathbf{p}} A_{\mathbf{p}\mathbf{p}'} x_{\mathbf{p}'} - i \sum_{\mathbf{p}} \epsilon_{\mathbf{p}} x_{\mathbf{p}}}  
\end{equation}
with $x_{\mathbf{p}} \in \mathbb{R}$ and $\mathbf{p}=(p_1,p_2)$, $p_1,p_2 \in [0,..,L-1]$. 
The linear part in the exponent, i.e. $\epsilon_{\mathbf{p}}$, is fixed by the static charge configuration (see section~\ref{modelsection} and Appendix~\ref{appformulation}) and 
\begin{equation}
A_{\mathbf{p}\mathbf{p}'} \equiv \frac{1}{\pi L^2 } \sum_{k_1,k_2=0}^{L-1} e^{2\pi i \frac{\left(p_1-p'_1\right) k_1 + \left(p_2-p'_2\right) k_2}{L}} \left(\gamma_{\mathbf{k}}^R+i \gamma_{\mathbf{k}}^I\right)
\end{equation}
is defined by the variational parameters $\left\{ \gamma_{\mathbf{k}}^R \right\} $ and $\left\{ \gamma_{\mathbf{k}}^I \right\}$. In the following, we will use the shorthand notation $\mathbf{p} \mathbf{k} \equiv 2 \pi \frac{p_1 k_1 + p_2 k_2}{L}$. Since the disorder introduced by static charges is incorporated in $\epsilon_{\mathbf{p}}$, the quadratic part $A$ is assumed to be translationally invariant.  
The factor of $1/\pi$ is chosen for later convenience. 

Written in terms of  Fourier components $x_{\mathbf{k}} = \frac{1}{L} \sum_{\mathbf{p}} e^{i \mathbf{p} \mathbf{k}} x_{\mathbf{p}}$, the quadratic part in the exponential becomes $\sum_{\mathbf{p},\mathbf{p'}} x_{\mathbf{p}} A_{\mathbf{p}\mathbf{p}'} x_{\mathbf{p}'} = \frac{1}{\pi} \sum_{\mathbf{k}} |x_{\mathbf{k}}|^2 \left(\gamma_{\mathbf{k}}^R+i \gamma_{\mathbf{k}}^I\right)$. 
Thus, to guarantee convergence of $\Psi_{CG}$ we need to require $\gamma_{\mathbf{k}}^R > 0 \hspace{2pt} \forall \mathbf{k}$. 
Since $|x_{\mathbf{k}}|^2=|x_{\mathbf{-k}}|^2$, the variational parameters $\gamma_{\mathbf{k}}^{R/I}$ and $\gamma_{-\mathbf{k}}^{R/I}$ are redundant. 
We define the equivalence relation
\begin{equation} \label{defkr}
    \begin{aligned}
        \mathbf{k} \sim_k \mathbf{k}' \qq{if} & k_1=-k'_1 \pmod{L} \\
        \text{and}\quad&k_2=-k'_2 \pmod{L}
    \end{aligned}
\end{equation}
With the quotient set $\mathcal{K} \equiv \left\{ [0,..,L-1]^2\setminus{(0,0)}\right\}/{\sim_k}$ we can define a set of independent variational parameters, $\left\{ \gamma^{R/I}_{\mathbf{k}}\right \}_{\mathbf{k} \in \mathcal{K}}$. 
Choosing a set of independent parameters will be important later on for applying the time dependent variational principle (see section~\ref{TDVP}).

To construct a suitable ansatz state for compact $U(1)$ gauge fields $(\theta_{\mathbf{p}} \in \left[ -\pi, \pi \right])$
we sum over complex Gaussian states, thus ensuring periodicity:
\begin{align}
\Psi_{CPG}\left(\{ \theta_{\mathbf{p}} \}\right)\equiv \prod_{\mathbf{p}} \left(\sum_{N_{\mathbf{p}}=-\infty}^{+\infty} \right) &\Psi_{CG}\left(\{ \theta_{\mathbf{p}} - 2\pi N_{\mathbf{p}} \}\right)\times \nonumber\\
& \times\delta \left(\sum_{\mathbf{p}} \theta_{\mathbf{p}} - 2\pi N_{\mathbf{p}}  \right).
\end{align}
The delta function needs to be included in order to satisfy condition~\eqref{globalconstraint} for physical states. 
To shorten notation, we will denote the product over infinite sums $\prod_{\mathbf{p}} \sum_{N_{\mathbf{p}}=-\infty}^{+\infty}$ by $\sum_{\{ N_{\mathbf{p}} \}}$ and the product over integrals $\prod_{\mathbf{p}} \int_{-\pi}^{\pi} d\theta_\mathbf{p}$ by $\int_{-\pi}^{\pi} D\theta$. 
The Gaussian nature of the wavefunction is exploited when evaluating expectation values of observables $O$ by combining the integral over $2\pi$ with one of the two infinite sums to an integration over the real axis
\begin{align} \label{formal_expectation_value}
\expval{O}{\Psi_{CPG}}=\sum_{\{N_{\mathbf{p}} \} } \delta\left({\sum_{\mathbf{p}} N_{\mathbf{p}}}\right) f_O(\{ N_{\mathbf{p}} \})
\end{align}
with
\begin{align}
&f_O(\{ N_{\mathbf{p}} \}) \nonumber\\
\equiv& \int\limits_{-\infty}^{+\infty} D\theta \hspace{1pt} \overline{\Psi_{CG}}\left({\theta_{\mathbf{p}}-2\pi N_{\mathbf{p}}}\right)  O\left({\theta_{\mathbf{p}}}\right) \Psi_{CG}\left({\theta_{\mathbf{p}}}\right) \delta\left({\sum_{\mathbf{p}} \theta_{\mathbf{p}}}\right).
\end{align}
The integral $f_O(\{ N_{\mathbf{p}} \})$ can be carried out analytically and the remaining infinite sum needs to be evaluated numerically. 

Exemplary, we show this procedure for the norm of the variational state, $\braket{\Psi_{CPG}}$. 
The computation of observables follows analagously; details on their exact form can be found in Appendix~\ref{observables}. 
After carrying out the integrals, the remaining function $f_1\left(\{ N_{\mathbf{p}} \}\right)$ is
\begin{equation}\label{partsum}
\begin{aligned}
f_1(\{ N_{\mathbf{p}} \})=\prod\limits_{\mathbf{k} \neq 0} \sqrt{\frac{\pi}{ \gamma_{\mathbf{k}}^R}} e^{2\pi i  \sum_{\mathbf{p}} \epsilon_{\mathbf{p}} N_{\mathbf{p}}}   e^{- \pi \sum_{\mathbf{k}} |N_{\mathbf{k}}|^2 \gamma_{\mathbf{k}} }
\end{aligned}      
\end{equation}
with $N_{\mathbf{k}} \equiv \frac{1}{L} \sum_{\mathbf{p}} e^{i \mathbf{p} \mathbf{k}} N_{\mathbf{p}}$ the discrete Fourier transform of $N_{\mathbf{p}}$ and $\gamma_{\mathbf{k}} \equiv \gamma_{\mathbf{k}}^R + (\gamma_{\mathbf{k}}^I)^2 (\gamma_{\mathbf{k}}^R)^{-1} $. The $\gamma_{\mathbf{k}}$ parameters determine how fast contributions to the sum in eq.~\eqref{formal_expectation_value} decrease exponentially with increasing $|N_{\mathbf{k}}|^2$. 

We group the configurations $N_{\mathbf{p}}$ of this sum in different orders such that within one order the configurations only change up to permutations. 
Since all relevant configurations will contain mostly zeros, we will denote orders by its non-zero elements, e.g. $\{ N \}_{1}$ is the set of all permutations of the configuration $N'$ defined by $N'_{\mathbf{p}=0}=1$ and $N'_{\mathbf{p} \neq 0}=0$, i.e. $\{ N \}_{1} \equiv S_{N'}$. 
If the parameters $\gamma_{\mathbf{k}}$ are large enough, the sum can be approximated by orders having small Euclidean norm, $||N_{\mathbf{p}} ||_{2}^2 = \sum_{\mathbf{p}} |N_{\mathbf{p}}|^2 = ||N_{\mathbf{k}}||^2_2$. 
The higher number of permutations in orders with larger norm cannot compensate for the exponential suppression (this would not be the case if the $\gamma_{\mathbf{k}}$ were arbitrarily small).
Using this scheme, the constraint $\delta\left(\sum_{\mathbf{p}} N_{\mathbf{p}}\right)$ is useful since it excludes many orders, e.g. $\{ N \}_{1}$ or $\{ N \}_{-1}$. 
The order with the lowest non-zero norm is therefore $\{ N \}_{1,-1}$. 
In fact, the sum in eq.~\eqref{formal_expectation_value} can be expanded in orders containing only pairs of $1,-1$:
\begin{align} 
&\braket{\Psi_{CPG}}\nonumber\\
=& \prod\limits_{\mathbf{k} \neq 0} \sqrt{\frac{\pi}{ \gamma_{\mathbf{k}}^R}} \sum_{ \{ N_{\mathbf{k}=\mathbf{0}}=0 \}} e^{2\pi i  \sum_{\mathbf{p}} \epsilon_{\mathbf{p}} N_{\mathbf{p}}}   e^{- \pi \sum_{\mathbf{k}} |N_{\mathbf{k}}|^2  \gamma_{\mathbf{k}} } \nonumber\\
=&\prod\limits_{\mathbf{k} \neq 0} \sqrt{\frac{\pi}{ \gamma_{\mathbf{k}}^R}} \left( 1 + \sum_{ \{ N \}_{1,-1}} e^{2\pi i  \sum_{\mathbf{p}} \epsilon_{\mathbf{p}} N_{\mathbf{p}}}   e^{- \pi \sum_{\mathbf{k}} |N_{\mathbf{k}}|^2  \gamma_{\mathbf{k}} }\right. \nonumber\\
&+\left. \sum_{ \{ N \}_{1,1,-1,-1}} e^{2\pi i \sum_{\mathbf{p}} \epsilon_{\mathbf{p}} N_{\mathbf{p}}}   e^{- \pi \sum_{\mathbf{k}} |N_{\mathbf{k}}|^2  \gamma_{\mathbf{k}} } + .. \right).
\label{expansion}
\end{align}
$\sum_{ \{ N_{\mathbf{k}=\mathbf{0}}=0 \}}$ denotes the sum over the set of all $N_{\mathbf{p}}$ configurations with $N_{\mathbf{k}=\mathbf{0}}=0$, i.e. fulfilling the global constraint. 
For sufficiently large $\gamma_{\mathbf{k}}$ higher orders of the type $\{ N \}_{2,-2}$ or $\{ N \}_{-2,1,1}$ are exponentially suppressed as well as orders with a large number of $1,-1$ pairs. 
Thus, the above expansion can be truncated after the first few terms. 
Each of the remaining orders is evaluated numerically. 
The fact that configurations only change up to permutations within one order can be used to highly parallelize the computation. 
On an $8 \times 8$ lattice we are able to compute the first three orders exactly. 
This procedure is sufficient for most configurations of variational parameters with $\gamma_{\mathbf{k}} \gtrsim 1$. 
However, in the intermediate regime $\gamma_{\mathbf{k}} \approx 1$ more orders are required to obtain good convergence. 
In these cases, higher orders are computed using uniform sampling. 
Since for all our purposes the different $\gamma_{\mathbf{k}}$ parameters were of the same order of magnitude and the $N_{\mathbf{p}}$ configurations only change up to a permutation within an order, a uniform probability distribution is a suitable ansatz for the exponential in eq.~\eqref{partsum}. 
This is only the case for sampling within one order; it would fail if one tried to sample the whole sum. 
This combined approach of exact evaluation and uniform sampling has the advantage that it introduces almost no error for most of the variational manifold  (up to truncated orders which are exponentially suppressed) and even for regions where uniform sampling is required the error is still suppressed since it only occurs in higher orders. For a detailed error analysis due to truncating orders and uniform sampling see Appendix~\ref{evaluation ansatz}. 

When the $\gamma_{\mathbf{k}}$ become small, the above approximation fails. 
In that case, one can exploit the fact that $\braket{\Psi_{CPG}}$ can be written as a multidimensional Riemann theta function~\cite{deconinck2003computing} which is defined as 
\begin{align}
\theta(z|\Omega)=\sum_{N \in \mathbb{Z}^g} e^{2\pi i (z \cdot N + \frac{1}{2} N \cdot \Omega \cdot N )}
\end{align}
where $z \in \mathbb{C}^g$, $\Omega \in \mathbb{C}^{g \times g}$, such that $\Omega=\Omega^{T}$ and $\mathrm{Im}(\Omega)$ is strictly positive definite. 
To bring $\braket{\Psi_{CPG}}$ into this form one can rewrite the delta function as the limit of a Gaussian and exchange the limit with the infinite sum due to uniform convergence. 
One can now exploit invariance of the Riemann theta function under modular transformations, in particular the following relation holds (for details see~\cite{deconinck2003computing}):
\begin{align}
\theta\left(z|\Omega\right)= \frac{1}{\sqrt{\det(-i \Omega)}} e^{-i \pi z \cdot \Omega \cdot z} \theta\left(\Omega^{-1} z | - \Omega^{-1}\right)
\end{align}
If we insert this relation and take the limit, we obtain: 
\begin{align}
\braket{\Psi_{CPG}}&=\prod\limits_{\mathbf{k} \neq 0} \sqrt{\frac{\pi}{ \gamma_{\mathbf{k}}^R \gamma_{\mathbf{k}}}} \sum_{ \{ N_{\mathbf{p}}  \}}   e^{- \pi  \sum_{\mathbf{k}} |N_{\mathbf{k}}-\epsilon_{\mathbf{k}}|^2  \gamma_{\mathbf{k}}^{-1}} \nonumber\\
&\equiv \sum_{ \{ N_{\mathbf{p}}  \}} f_{\mathrm{inv},1}\left(\{ N_{\mathbf{k} \neq 0} \}\right).
\end{align} 
with $\gamma_{\mathbf{0}}^{-1}=0$. The exponential weight depends now on $\gamma_{\mathbf{k}}^{-1}$ which allows in principle to approximate the sum with only a very limited number of orders for sufficiently small $\gamma_{\mathbf{k}}$. 
However, the sum is not well defined since all constant configurations $N_{\mathbf{p}}=c (1,1,...,1)$ have weight one for $c \in \mathbb{Z}$. 
Fortunately, since all $f_{\mathrm{inv},O}(\{ N_{\mathbf{k} \neq 0} \})$ are independent of $N_{\mathbf{k}=\mathbf{0}}$ (as a result of the global constraint on physical states), all these configurations can be factored out such that they cancel when calculating expectation values. 
This can be formulated rigorously by defining an equivalence relation for $N_{\mathbf{p}}$ configurations:
\begin{align}
N_{\mathbf{p}} \sim_1 N_{\mathbf{p}}' \hspace{10pt} \mathrm{if} \hspace{5pt} \exists \hspace{10pt}  c \in \mathbb{Z} \hspace{10pt} \mathrm{s.t.} \hspace{5pt} N_{\mathbf{p}}-N_{\mathbf{p}}'= c (1,1,...,1) 
\end{align}
When calculating expectation values of observables only a sum over representatives of this equivalence relation is required:
\begin{align}
\frac{\expval{O}{\Psi_{CPG}}}{\braket{\Psi_{CPG}}}= \frac{\sum_{ \{ N_{\mathbf{p}}  \}/\sim_1 } f_{\mathrm{inv},O}(\{ N_{\mathbf{k} \neq 0} \})}{\sum_{ \{ N_{\mathbf{p}}  \}/\sim_1} f_{\mathrm{inv},1}(\{ N_{\mathbf{k} \neq 0} \})}
\end{align}
If we choose the representative to be the one closest in norm to the $N_{\mathbf{p}}=\mathbf{0}$ configuration, we can expand the sum again in orders having mostly $0$'s. In this case we have no constraint so that all orders must be taken into account. 
For more details see Appendix~\ref{evaluation ansatz}. 

A nice way to check the validity of both numerical approximation schemes presented above is to see whether they agree in the parameter region $\gamma_{\mathbf{k}} \approx 1 $. 
This check has been carried out throughout this work since it also indicates that the whole variational manifold can be accessed which is required in order to study the whole coupling region. 

To illustrate that both approximation schemes complement each other, we give the variational energy of $\Psi_{CPG}$ with respect to the Kogut-Susskind Hamiltonian given in eq.~\eqref{Hamiltonianref}, written both in the infinite sum representation for high and for low $\gamma_{\mathbf{k}}$: 
\begin{widetext}
\begin{equation} \label{varenergy}
\begin{aligned}
\frac{ \expval{H_{KS}}{\Psi_{CPG}}}{\braket{\Psi_{CPG}}} =& E_C+\frac{g^2}{4\pi} \sum_{\mathbf{k}} \gamma_{\mathbf{k}} \left(4-2 \cos\left(\frac{2\pi k_{1}}{L}\right)-2 \cos\left(\frac{2\pi k_{2}}{L}\right) \right)\\
&-\frac{g^2}{2} \sum_{\mathbf{k}} \gamma_{\mathbf{k}}^2 \left(4-2 \cos\left(\frac{2\pi k_{1}}{L}\right)-2 \cos\left(\frac{2\pi k_{2}}{L}\right) \right) \expval{|N_{\mathbf{k}}|^2}\\
&+\frac{1}{g^2}  \sum_{\mathbf{p}} \left( 1-  e^{-\frac{\pi}{4L^2} \sum_{\mathbf{k} \neq \mathbf{0}} \left(\gamma_{\mathbf{k}}^R\right)^{-1}} \expval{(-1)^{N_{\mathbf{p}}} \cosh\left(\pi \sum_{\mathbf{k}} \mathrm{Re}\left(N_{\mathbf{k}} b_{\mathbf{k}}^{\mathbf{p}}\right)\right)   }  \right)
\end{aligned}
\end{equation}
with $b_{\mathbf{k}}^{\mathbf{p}} =  \frac{1}{L} \gamma_{\mathbf{k}}^I\left(\gamma_{\mathbf{k}}^R\right)^{-1} e^{-i\mathbf{p}\mathbf{k}}$. 
The brackets denote the infinite sums:
\begin{align}
\label{elinfinitesum} \expval{|N_{\mathbf{k}}|^2} \equiv&    \frac{\sum_{ \{ N_{\mathbf{k}=\mathbf{0}}=0 \}} e^{2\pi i \sum_{\mathbf{p}} \epsilon_{\mathbf{p}} N_{\mathbf{p}}}   e^{- \pi \sum_{\mathbf{k'}} |N_{\mathbf{k'}}|^2  \gamma_{\mathbf{k'}} } |N_{\mathbf{k}}|^2}{\sum_{ \{ N_{\mathbf{k}=\mathbf{0}}=0 \}} e^{2\pi i \sum_{\mathbf{p}} \epsilon_{\mathbf{p}} N_{\mathbf{p}}}   e^{- \pi \sum_{\mathbf{k'}} |N_{\mathbf{k'}}|^2  \gamma_{\mathbf{k'}} }} \nonumber\\
=&\frac{1}{2\pi} \gamma_{\mathbf{k}}^{-1} \left(4-2 \cos\left(\frac{2\pi k_{1}}{L}\right)-2 \cos\left(\frac{2\pi k_{2}}{L}\right) \right) \nonumber\\
&- \gamma_{\mathbf{k}}^{-2} \frac{\sum_{ \{ N_{\mathbf{p}}  \}/\sim_1 } e^{- \pi \sum_{\mathbf{k'}} |N_{\mathbf{k'}} - \epsilon_{\mathbf{k'}}|^2 \gamma_{\mathbf{k'}}^{-1} } |N_{\mathbf{k}}-\epsilon_{\mathbf{k}}|^2}{\sum_{ \{ N_{\mathbf{p}}  \}/\sim_1 } e^{- \pi \sum_{\mathbf{k'}}|N_{\mathbf{k'}} - \epsilon_{\mathbf{k'}}|^2 \gamma_{\mathbf{k'}}^{-1} }} \\
\label{maginfinitesum} \expval{(-1)^{N_{\mathbf{p}}} \cosh(\pi \sum_{\mathbf{k}} \mathrm{Re}\left(N_{\mathbf{k}} b_{\mathbf{k}}^{\mathbf{p}}\right))    }=& \frac{\sum_{ \{ N_{\mathbf{k}=\mathbf{0}}=0 \}} (-1)^{N_{\mathbf{p}}}  \cosh\left(\pi \sum_{\mathbf{k}} \mathrm{Re}\left(N_{\mathbf{k}} b_{\mathbf{k}}^{\mathbf{p}}\right)\right)   e^{2\pi i  \sum_{\mathbf{p}} \epsilon_{\mathbf{p}} N_{\mathbf{p}}}   e^{- \pi \sum_{\mathbf{k}} |N_{\mathbf{k}}|^2  \gamma_{\mathbf{k}} }  }{\sum_{ \{ N_{\mathbf{k}=\mathbf{0}}=0 \}} e^{2\pi i \sum_{\mathbf{p}} \epsilon_{\mathbf{p}} N_{\mathbf{p}}}   e^{- \pi \sum_{\mathbf{k}} |N_{\mathbf{k}}|^2  \gamma_{\mathbf{k}} }}   \nonumber\\
=& \frac{\sum_{ \{ N_{\mathbf{p}}  \}/\sim_1 } e^{-\pi \sum_{\mathbf{k}} \big(|N_\mathbf{k} - \epsilon_{\mathbf{k}} - \frac{1}{2}^{\mathbf{p}}_\mathbf{k}|^2 -\frac{1}{4} |b_{\mathbf{k}}^{\mathbf{p}}|^2 \big) \gamma_{\mathbf{k}}^{-1} } \cos \left( \pi \sum_\mathbf{k} \gamma_{\mathbf{k}}^{-1} \mathrm{Re} \left[ \big(N_\mathbf{k} - \epsilon_\mathbf{k} - \frac{1}{2}^{\mathbf{p}}_\mathbf{k}\big) b_{\mathbf{k}}^{\mathbf{p}} \right] \right)}{\sum_{ \{ N_{\mathbf{p}}  \}/\sim_1 } e^{- \pi \sum_{\mathbf{k}} |N_{\mathbf{k}} - \epsilon_{\mathbf{k}}|^2 \gamma_{\mathbf{k}}^{-1} } }
\end{align}
\end{widetext}
with $\frac{1}{2}^{\mathbf{p}}_\mathbf{k}= \frac{1}{2L} e^{-i\mathbf{p}\mathbf{k}}$. 
If we set $\gamma_{\mathbf{k}}^I = 0 \hspace{3pt} \forall \hspace{1pt} \mathbf{k}$ the expressions for high $\gamma_{\mathbf{k}}^R$, i.e. with the sums $\sum_{ \{ N_{\mathbf{k}=\mathbf{0}}=0 \}}$, agree with the results given in~\cite{drell1979quantum} up to redefinitions. 
It is important to emphasize that the convergence of infinite sums is determined by $\gamma_{\mathbf{k}} = \gamma_{\mathbf{k}}^R + \left(\gamma_{\mathbf{k}}^I\right)^2 \left(\gamma_{\mathbf{k}}^R\right)^{-1}$  or $\gamma_{\mathbf{k}}^{-1}$, respectively. 
For real-time evolutions, e.g. a quantum quench, $\left(\gamma_{\mathbf{k}}^I\right)^2$ will typically become large and so will $\gamma_{\mathbf{k}}$, irrespective of the real part $\gamma_{\mathbf{k}}^R$. 
This allows to truncate the expansion in eq.~\eqref{expansion} already after the first term such that everything can be evaluated without resorting to sampling. 
This property makes the ansatz well suited for real-time evolution compared with other methods where sampling at all times often makes it difficult to reach long times.

\section{\label{static} Static properties}

In this section, we study the variational ground state of 2+1d compact QED over the whole coupling region. 
To minimize the energy we applied a gradient descent algorithm (the formula for the gradient can be found in Appendix~\ref{observables}). 
We used different initial seeds to prevent the possibility of getting stuck in local minima. 
To make sure that our variational state can approximate the ground state, we compare it first to known exact results. 
One should note that exact diagonalization methods cannot be applied to the full theory since the local Hilbert space is infinite. 
However, for the case of a single plaquette exact analytical solutions are known, namely the Mathieu functions. 

\subsection{Benchmark for one plaquette}
\begin{figure}
	\centering
	\includegraphics[width=\columnwidth]{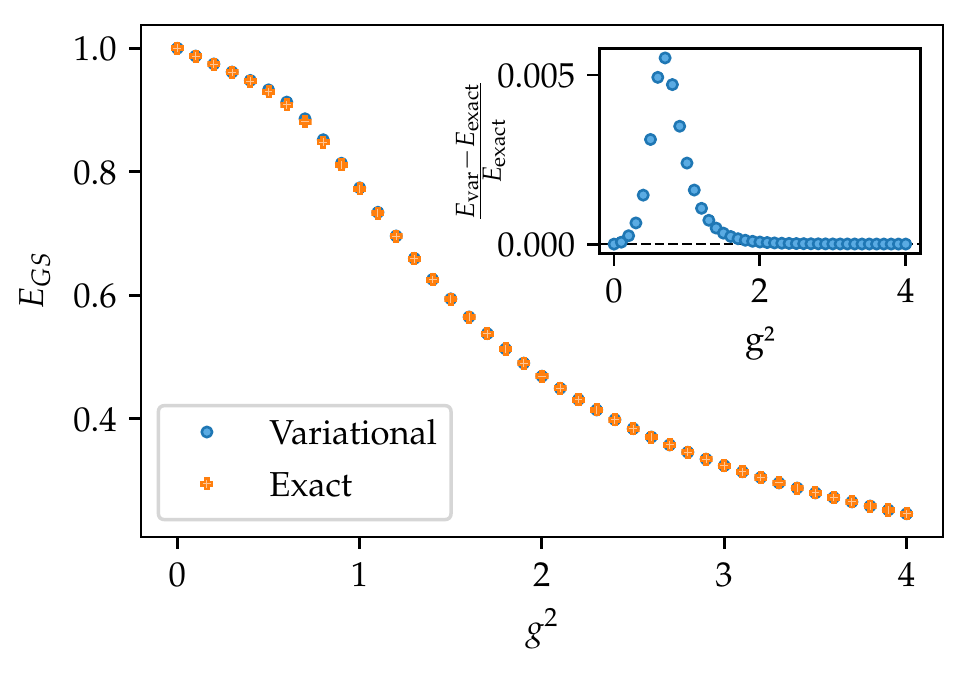}
	\caption{Benchmark of the variational ground state energy for one plaquette against the value of the exact ground state, given by the Mathieu function with the lowest characteristic value. The inset shows the relative error of the variational ground state energy with respect to the exact goundstate energy.}
	\label{Benchmark_one_plaquette}
\end{figure}

For benchmarking our variational ansatz, we will restrict ourselves to the sector without static charges. The Hamiltonian given in the formulation of the previous chapter, written in the basis of $\theta$, reads: 

\begin{align}
	H_{1 \mathrm{plaq}}= -2g^2 \frac{\partial^2}{\partial \theta^2} + \frac{1}{g^2} (1-\cos \theta).
\end{align} 
The corresponding Schroedinger equation for $\xi(\theta)$ can be written as a Mathieu equation: 

\begin{equation}
\left( \frac{\partial^2}{\partial z^2}+a -2q \cos(2z) \right) \Tilde{\xi}(z) = 0
\end{equation}
with $q \equiv -\frac{1}{g^4}$, $a \equiv \frac{2}{g^2}\left( E - \frac{1}{g^2} \right)$ and $\Tilde{\xi}(z) \equiv \xi(\theta /2)$. 
$\Tilde{\xi}$ is therefore not $2 \pi$-periodic but $\pi$-periodic. 
The $\pi$-periodic solutions are usually separated into even $ce_{2r}(z,q) \hspace{2pt} (r \geq 0)$ and odd $se_{2r}(z,q) \hspace{2pt}  (r \geq 1)$ solutions. 
The lowest energy, i.e. the lowest characteristic value $a$, corresponds to the solution $ce_{0} (z,q)$. 
In Fig.~\ref{Benchmark_one_plaquette}, this exact ground state energy is plotted against the minimized variational energy. 
They agree very well over the whole coupling region, even in the regime where the difference is maximal ($g^2 \sim 0.7 $) the relative error is still around $0.5 \% $. 

\subsection{Ground state properties}
\begin{figure}
	\centering
	\includegraphics[width=\columnwidth]{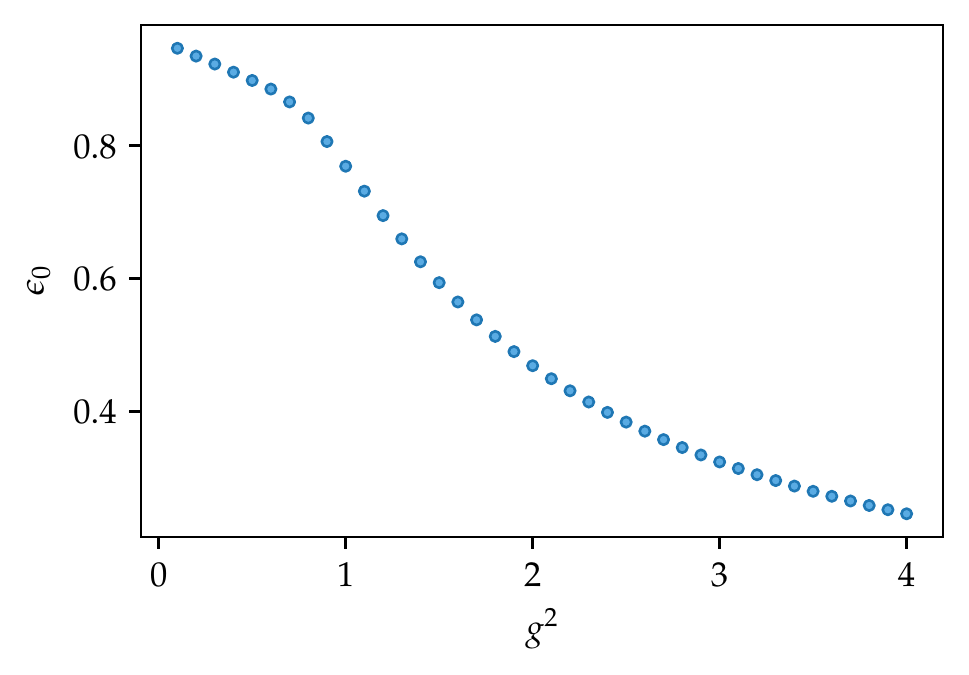}
	\caption{Groundstate energy density extrapolated to the thermodynamic limit. The available lattice sizes are $8 \times 8$ for couplings $g^2 \geq 1.0$, $14 \times 14$ for $g^2=0.8,0.9$ and $20 \times 20$ for $g^2 \leq 0.7$. }
	\label{GS_thermo}
\end{figure}

\begin{figure} \label{finite size}
	\centering
	\includegraphics[width=\columnwidth]{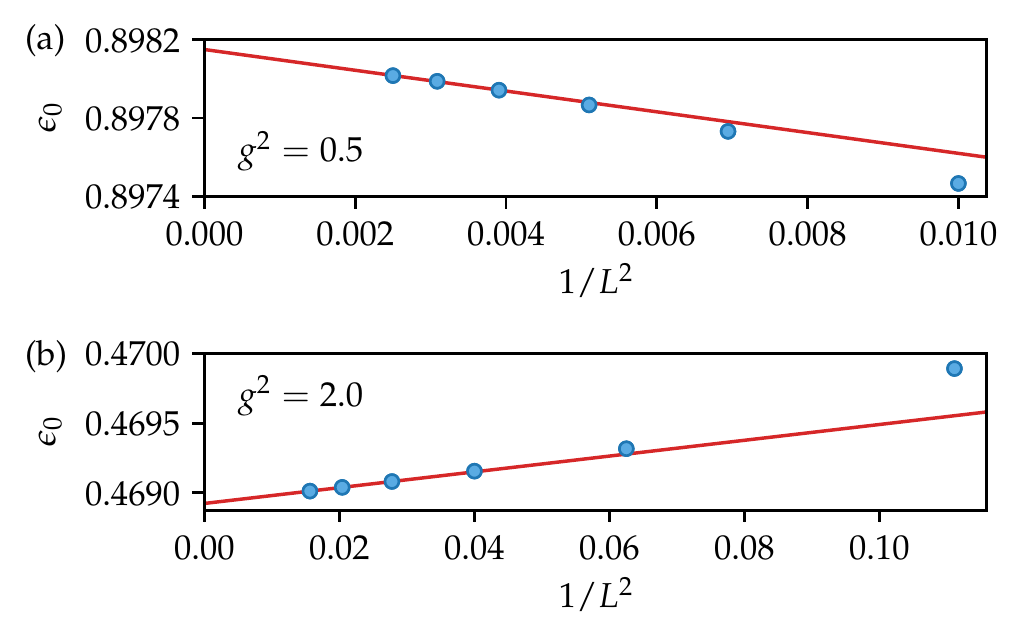}
	\caption{Finite size scaling for the ground state energy density at $g^2=0.5$~(a) and $g^2=2.0$~(b). For $g^2=2.0$, the ground state energy density for $L=8,7,6$ is fitted according to eq.~\eqref{finitesizeformula}. The remaining data points correspond to $L=5,4,3$. For $g^2=0.5$, lattice sizes of $L=20,18,16$ are used for the fit, the remaining data points correspond to $L=14,12,10$. }
	\label{finitesize}
\end{figure}

In this section, we study the properties of the varational ground state for an extended lattice and investigate its finite size effects. 
We start by studying the ground state energy density $e_{0}(L)$ for lattice sizes up to $8 \times 8$ plaquettes without static charges. 
We see that for couplings $g^2 \gtrsim 1.0 $ this size is already enough to get a linear scaling with $\frac{1}{L^2}$. The thermodynamic limit $e_{0}(L=\infty)$ is then extracted with the following fit
\begin{equation} \label{finitesizeformula}
e_{0}(L) = e_{0}(L=\infty) + \frac{a}{L^2}.
\end{equation}
For large couplings the thermodynamic limit can be reached with even smaller lattice sizes. 
The region which limits the evaluation of our variational state to $8  \times 8$ is around $g^2 \sim 1.1$ since the variational parameters are of order one ($\gamma^R_{\mathbf{k}} \sim 1 $ , $\gamma^I_{\mathbf{k}}=0 $) and thus both approximation schemes agree (see Appendix~\ref{evaluation ansatz}). 
Hence, for couplings below this transition region we can simulate larger lattices, namely $14 \times 14$ for $g^2=0.8,0.9$ and $20 \times 20$ for $0.1 \leq g^2 \leq 0.7$. 
For such lattice sizes, the finite size effects become again small enough to extrapolate to the thermodynamic limit. 
The result for the ground state energy density in the thermodynamic limit over the whole coupling region is shown in Fig.~\ref{GS_thermo}. To illustrate, we show the extrapolation to the thermodynamic limit for $g^2=0.5$ and $g^2=2.0$ in Fig.~\ref{finitesize}.

\begin{figure}
	\centering
	\includegraphics[width=\columnwidth]{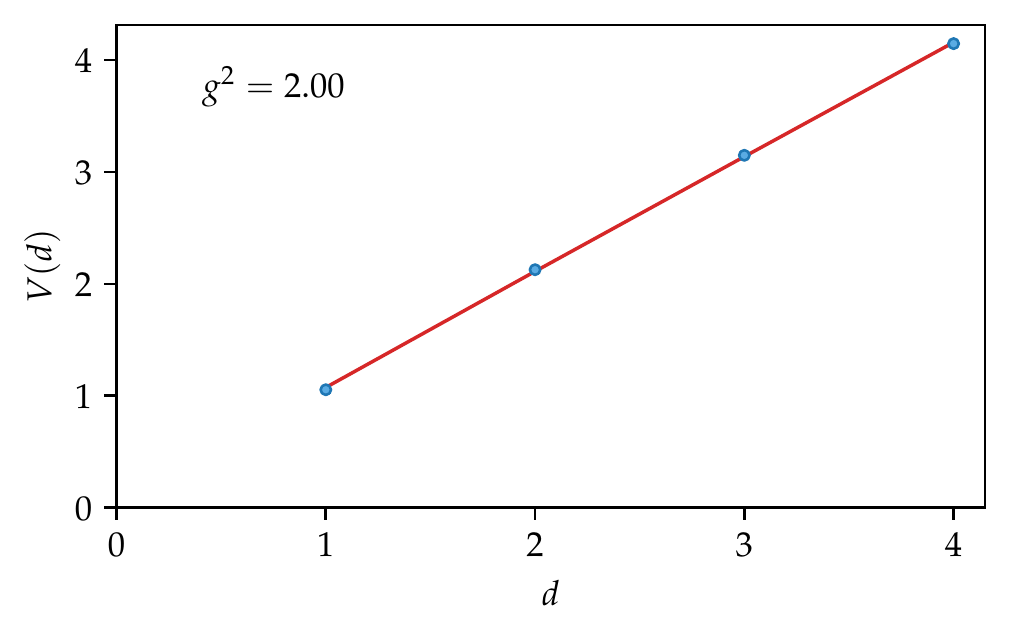}
	\caption{The static potential $V(d)$ of two charges separated by a distance $d$ at $g^2=2.0$. The data points are computed on an $8 \times 8$ lattice as the difference between the ground state energy with the respective static charge configuration and the ground state energy without static charges. The red line is a fit to the potential according to eq.~\eqref{staticpotentialformula} with $\sigma=1.001$ and $b=0.146$.}
	\label{static_potential_fit}
\end{figure}

In the next step, we study the string tension over the whole coupling region. 
We can measure it in two ways: First, we place static charges and analyze the scaling of the ground state energy depending on the distance between static charges. 
We will fit the potential with the following function: 

\begin{align} \label{staticpotentialformula}
V(d)= \sigma d + b V_{Coul}(d)  
\end{align} 
where $\sigma$ is the string tension and $V_{Coul}$ is the lattice Coulomb potential in two dimensions which becomes a logarithmic potential in the continuum limit. The values for $V(d)$ are computed as the difference between the ground state energy with static charges separated by a distance $d$ and the ground state energy without static charges. Exemplary, we show the fit of the potential for $g^2=2.0$ in Fig.~\ref{static_potential_fit}.

\begin{figure}
	\centering
	\includegraphics[width=\columnwidth]{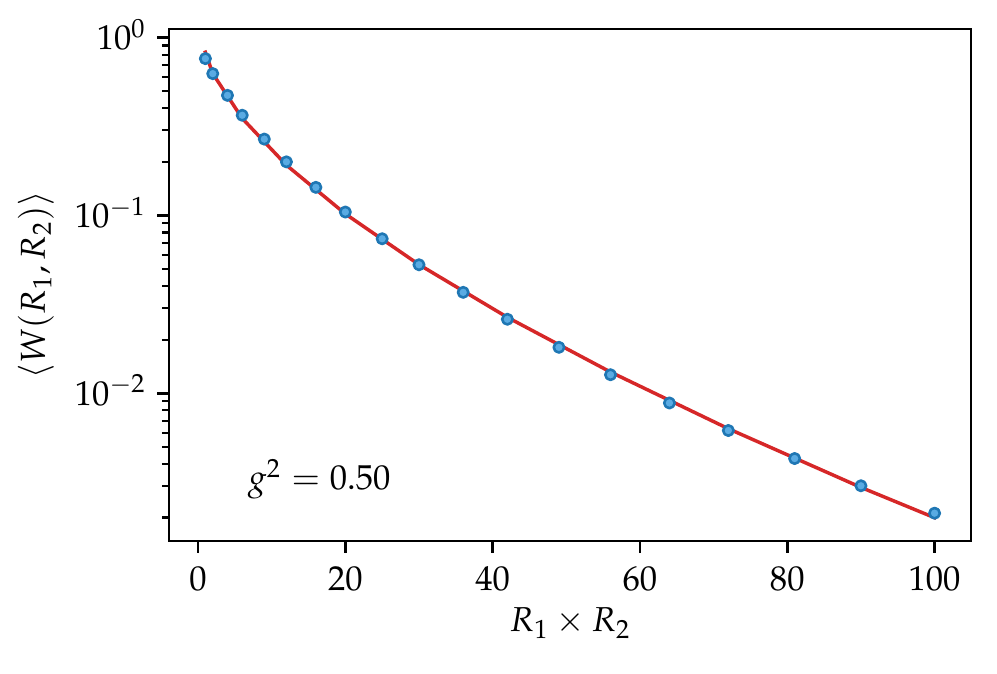}
	\caption{The data points show different spatial Wilson loops $\expval{W(R_1,R_2)}$ in the ground state at $g^2=0.5$, computed on a $20 \times 20$ lattice, as a function of the area $R_1 \times R_2$. The maximally used edge length of a Wilson loops is $10$ ($R_1,R_2 \leq 10$), with a maximum difference between the edges of one ($|R_2-R_1| \leq 1$). The red line is a fit to the exponential decay of Wilson loops according to eq.~\eqref{wilsonloopformula} with $\sigma=0.013$, $a=0.132$ and $c=0.349$.}
	\label{wilson_fit}
\end{figure}
In the second approach we use the scaling of spatial Wilson loops to extract the string tension. 
This works at zero temperature since on the Euclidean lattice spatial and temporal Wilson loops are related by $O(4)$ symmetry. At finite temperature this symmetry is broken due to a  compactified temporal dimension \cite{svetitsky1982critical}. The formula to calculate Wilson loops of arbitrary size with complex periodic Gaussian states in both the low and high $\gamma_{\mathbf{k}}$ approximation can be found in Appendix~\ref{observables}. 
On $8 \times 8$ lattices, we consider all rectangular loops $R_1 \times R_2$ with $R_1,R_2 \leq 4$  (four is the maximal physical length due to the periodic boundary conditions). 
Furthermore, we require $|R_1 - R_2| \leq 1$ to avoid additional finite size effects coming from an asymmetry in the edges. 
For weak couplings, where larger lattices are accessible, we extend the allowed maximal edge length to $7$ and $10$ (for $14 \times 14$, resp. $20 \times 20$). 
We fit the Wilson loop scaling according to the following formula: 

\begin{align}\label{wilsonloopformula}
W(R_1,R_2)= e^{-\sigma R_1 R_2 - 2a (R_1+R_2)  + c}
\end{align}
The first term corresponds to area law scaling with string tension $\sigma$ and the second term to perimeter law scaling. 
To illustrate the procedure, we show the fit for the ground state at $g^2=0.5$ in Fig.~\ref{wilson_fit}.
We also tried to extract the string tension via Creutz ratios \cite{creutzasymptoticfreedom1980} but the results were less reliable than the Wilson loop fits. 

\begin{figure}
	\centering
	\includegraphics[width=\columnwidth]{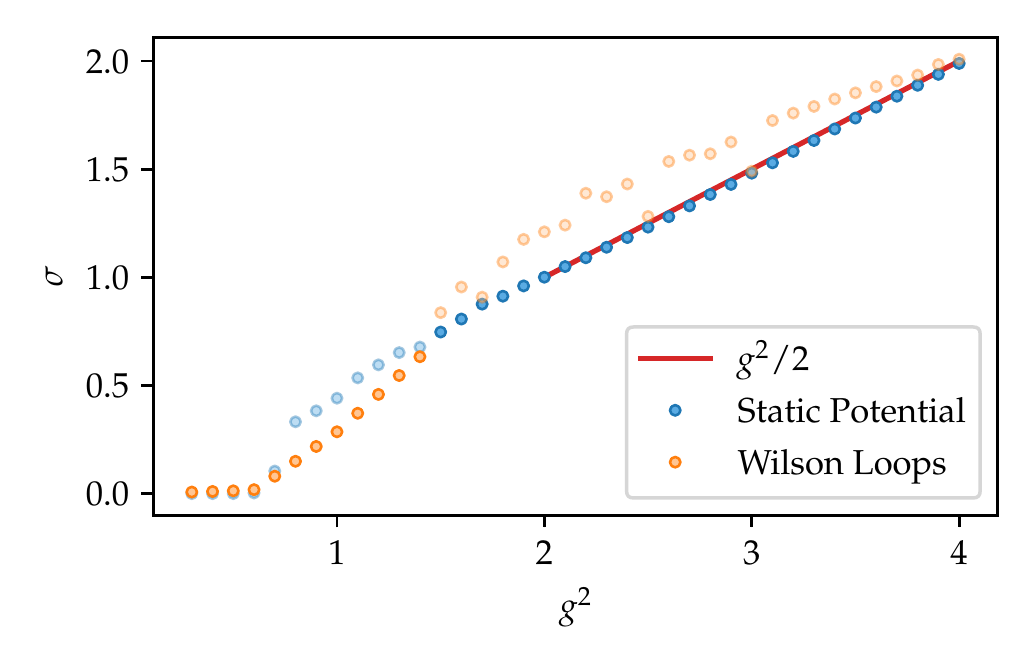}
	\caption{String tension fitted via the static potential (blue) and via the decay of spatial Wilson loops (orange). For larger couplings ($g^2 \geq 1.5$) the static potential fit performs better than the fit of Wilson loops and agrees with the strong-coupling prediction $g^2/2$. For small couplings ($g^2 \leq 1.4$) Wilson loop fits are more suitable. The more reliable method is shown with full data points while data points of the other method are made transparent. }
	\label{stringtension}
\end{figure}
The result for both approaches is shown in Fig.~\ref{stringtension}. 
For large values of the coupling constant, the fit for the static potential works well and agrees with the strong-coupling prediction $\frac{g^2}{2}$. 
Since a large coupling implies a significant distance from the continuum limit, moderate lattice sizes are sufficient to observe the onset of the linear part of the potential. The scaling of Wilson loops is prone to errors in that regime as expectation values of large Wilson loops become close to machine precision. 
However, for small couplings the Wilson loop scaling is the better method since expectation values of Wilson loops do not decay as fast due to the small string tension. 
Since both methods complement each other we chose to make the string tension data for the static potential transparent for couplings $g^2 \leq 1.5$ and the ones extracted by Wilson loops scaling for $g^2 > 1.5$. 
The remaining full data points in Fig.~\ref{stringtension} are the most reliable estimates for the string tension.

\begin{figure}
	\centering
	\includegraphics[width=\columnwidth]{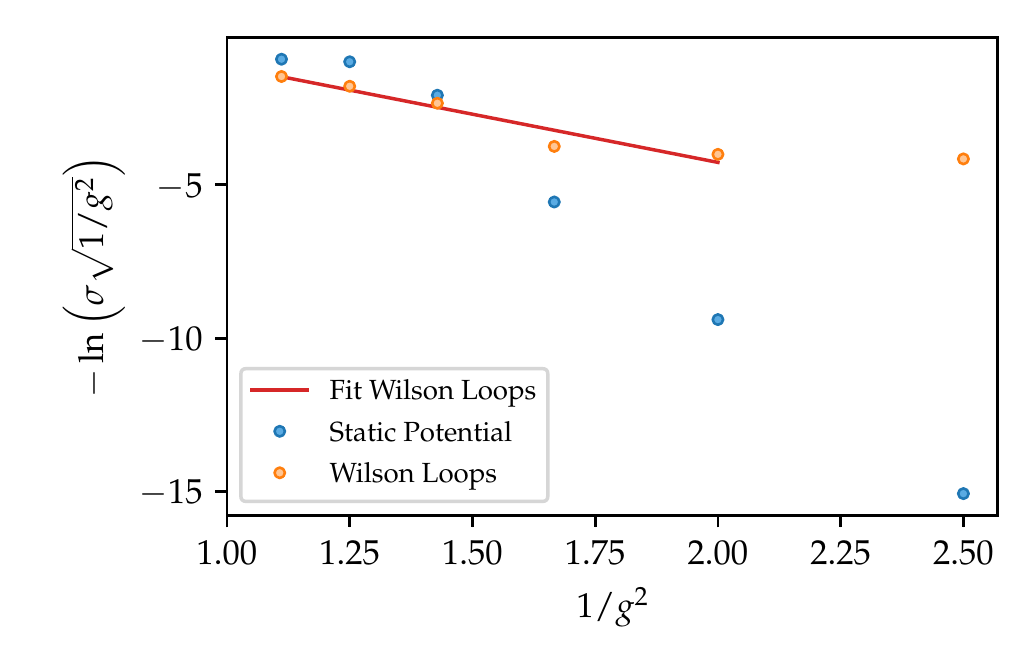}
	\caption{String tension in the weak-coupling regime. While the Wilson loop fits show exponential decay of the string tension close to the theoretical value ($\nu_0=0.318$ compared to $\nu_{0,\mathrm{theo}}=0.321$), the static potential fits become unreliable for couplings $g^2 \leq 0.6$.}
	\label{stringtensionexp}
\end{figure}

For small couplings an exponential decay of the string tension is expected according to the formula \cite{ambjorn1982string}: 

\begin{align}
\sigma= c \sqrt{\frac{g^2}{\pi^2}} e^{-\frac{\pi^2}{g^2} \nu_0}.
\end{align}
If we fit this formula to the string tension data of the Wilson loop fits between $0.5 \leq g^2 \leq 0.9$ (see Fig.~\ref{stringtensionexp}) we obtain $c=23.53$ and $\nu_0=0.318$ which is close to the theoretical prediction ($\nu_{0,\mathrm{theo}}=0.321$)~\cite{loan2003path}.

\subsection{Truncation effects}
\begin{figure}
	\centering
	\includegraphics[width=\columnwidth]{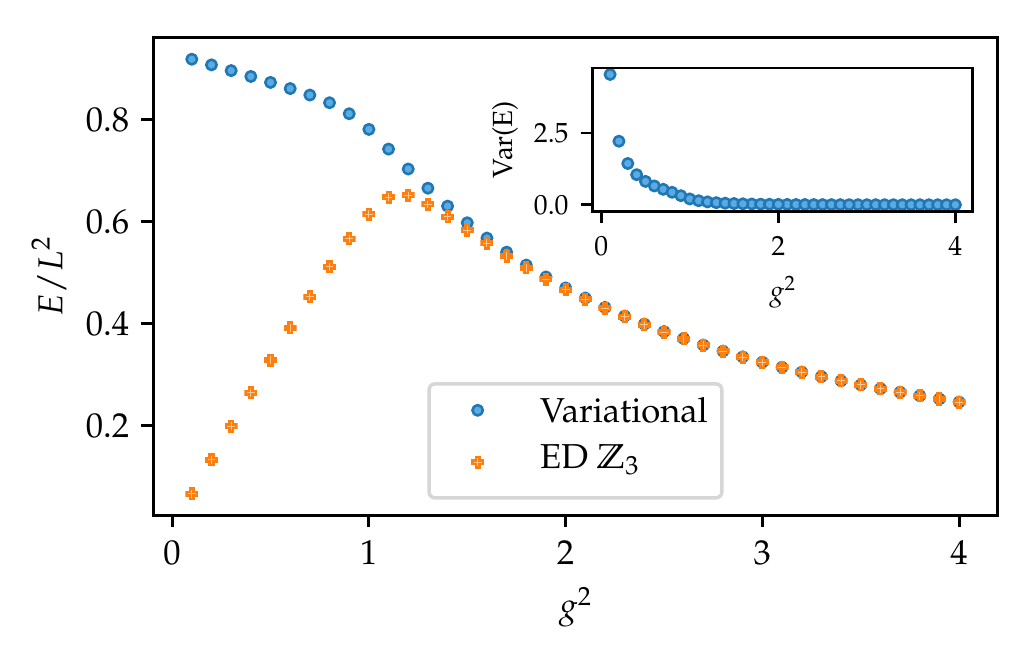}
	\caption{
    	Comparison of the ground state energy density on a $3 \times 3$ lattice without static charges, computed for a $\mathbb{Z}_3$ lattice gauge theory by exact diagonalization (orange) and for the full $U(1)$ theory by minimizing the variational energy (blue). The inset shows the variance of the electric field on a link in the variational ground states.
    }
	\label{comparison ED}
\end{figure}

Since our wave function does not require a truncation, we can study truncation effects of other methods.
Here, we will focus on a truncation in the electric basis. 
To see these effects we will study the variance of the electric field operator. 
For simplicity, we will look at this effect without static charges, since they only introduce $\epsilon$-shifts ($-1/2 < \epsilon < 1/2$) in the electric field. 
Since the expectation value of the electric field vanishes in the absence of static charges, we can write the variance in terms of the electric energy

\begin{align}
 \mathrm{Var} (E_{\mathbf{x},i}) &= \expval{E_{\mathbf{x},i}^2}-\expval{E_{\mathbf{x},i}}^2=\frac{1}{L^2 g^2} \expval{H_{E}}.
\end{align}
The variance is plotted in the inset of Fig.~\ref{comparison ED} for the ground state which was computed in the last section. To quantitatively show the difference, we compare our variational state to an exact diagonalization calculation of a $\mathbb{Z}_3$ lattice gauge theory. 
To reduce the required Hilbert space dimension, we formulate it in terms of plaquette variables, in the same style as we did for the $U(1)$ theory. 
The Hilbert space is truncated in the eigenbasis of $L_{\mathbf{p}}$ to three states (corresponding to the eigenvalues $m=0,1,-1$). 
To make this a consistent theory we define the gauge field operators cyclically: 

\begin{align}
U_p^{\dagger} \ket{m} = \ket{m'}\qq{with $m'=m+1 \pmod{3}$.}
\end{align}
This is equivalent to a $\mathbb{Z}_3$ lattice gauge theory formulated in link variables: 

\begin{align}
H_{Z3}= \frac{g^2}{6} \sum_{\mathbf{x},i} (2-P_{\mathbf{x},i}-P^{\dagger}_{\mathbf{x},i}) + \frac{1}{2g^2} \sum_{\mathbf{p}} (2- Q_{\mathbf{p}}-Q_{\mathbf{p}}^\dagger )
\end{align}
with $Q_{\mathbf{p}} \equiv Q_{\mathbf{x},1}Q_{\mathbf{x}+\mathbf{e}_{1},2} Q_{\mathbf{x}+\mathbf{e}_{2},1}^{\dagger}Q_{\mathbf{x},2}^{\dagger}$ where $\mathbf{x}$ is the vertex at the bottom left corner of plaquette $\mathbf{p}$ and $Q_{\mathbf{x},i}$ the cyclic raising operator of the electric field on link $(\mathbf{x},i)$, such that (see \cite{horn1979hamiltonian} for details)
\begin{align}
  &P_{\mathbf{x},i}^N=Q_{\mathbf{x},i}^N=1&& P_{\mathbf{x},i}^\dagger P_{\mathbf{x},i}=Q_{\mathbf{x},i}^\dagger Q_{\mathbf{x},i}=1\nonumber\\
  &P_{\mathbf{x},i}^\dagger Q_{\mathbf{x},i} P_{\mathbf{x},i}  =e^{i\frac{2\pi}{3}Q_{\mathbf{x},i}}. && 
  \label{eq:ZN_algebra}
\end{align}
The maximal lattice size we can achieve in our ED calculation for a reasonable amount of time is $3 \times 3$ plaquettes. 
We calculate the ground state energy density for this lattice size with ED and our variational ansatz. 
The result is shown in Fig.~\ref{comparison ED}.  The two approaches exhibit good agreement in the strong coupling regime. For intermediate couplings differences becomes more pronounced leading to  qualitatively different results in the weak-coupling limit $g \to 0$ . 

Since the electric Hamiltonian becomes bounded in the truncated theory, it does not contribute in the weak coupling limit. In the $U(1)$ theory, however, the electric Hamiltonian is unbounded and the growth in  electric energy leads to a finite result for the ground state energy in the continuum limit. 

\section{\label{dynamic} Real-time dynamics}
In this section, we study out-of-equilibrium dynamics by applying the following quench protocol: We prepare the ground state for the compact QED Hamiltonian at some coupling $g^2$, quench to a Hamiltonian with a different coupling constant $g^2_{\mathrm{quench}}$ and observe the subsequent time evolution. The observables we track during the evolution are Wilson loops and the electric field (their expectation values in terms of the variational parameters can be found in Appendix~\ref{observables}). In addition we check whether the energy is conserved throughout the whole time evolution.

\subsection{Time-dependent variational principle} \label{TDVP}
To study dynamical phenomena, we employ the time-dependent variational principle. 
The equations of motion are projected onto the tangent plane of our variational manifold. For every variational parameter $\gamma_{\mathbf{k}}^{R/I}$ we define a corresponding tangent vector $\ket{\Psi_{\mathbf{k}}^{R/I}} \equiv \mathbb{P}_{\Psi} \left( \frac{\partial}{\partial \gamma_{\mathbf{k}}^{R/I}} \ket{\Psi_{CPG}} \right)$ where $\mathbb{P}_{\Psi}$ ensures orthogonality to $\ket{\Psi_{CPG}}$: 

\begin{align} \label{projectionvarmanifold}
    \mathbb{P}_{\Psi}(\ket{\psi}) \equiv \ket{\psi}-\braket{\Psi_{CPG}}{\psi}\ket{\Psi_{CPG}}
\end{align}
If we restrict the momenta $\mathbf{k}$ of the variational parameters to the set $\mathcal{K}$ defined in eq.~\eqref{defkr}, all tangent vectors become linearly independent. This allows to invert the Gram matrix $G_{\mathbf{k'}\mathbf{k}} \equiv \braket{\Psi_{\mathbf{k'}}^R}{\Psi_{\mathbf{k}}^R}$ with $\mathbf{k},\mathbf{k'} \in \mathcal{K}$. Since our variational manifold is Kähler, we can express the time evolution of the variational parameters $\gamma_{\mathbf{k}}^{R/I}$ ($\mathbf{k} \in \mathcal{K}$) in the following way \cite{hackl2020geometry}:

\begin{align}
i \left(\dot{\gamma}_{\mathbf{k}}^R+i \dot{\gamma}_{\mathbf{k}}^I\right)&= \frac{1}{2} \sum_{\mathbf{k'} \in \mathcal{K}} (G^{-1})_{\mathbf{k}\mathbf{k'}}  \left( \frac{\partial E}{\partial \gamma_{\mathbf{k'}}^{R}} + i \frac{\partial E}{\partial \gamma_{\mathbf{k'}}^I} \right)  
\end{align}
with $E \equiv \frac{ \expval{H_{KS}}{\Psi_{CPG}}}{\braket{\Psi_{CPG}}}$ the variational energy in eq.~\eqref{varenergy} and $\dot{\gamma}\equiv\pdv{\gamma}{t}$. The formula for the calculation of the Gram matrix and the gradient of the variational energy can be found in Appendix~\ref{observables}. 

\subsection{Benchmark of variational ansatz} 

\begin{figure}
\centering
\includegraphics[width=\columnwidth]{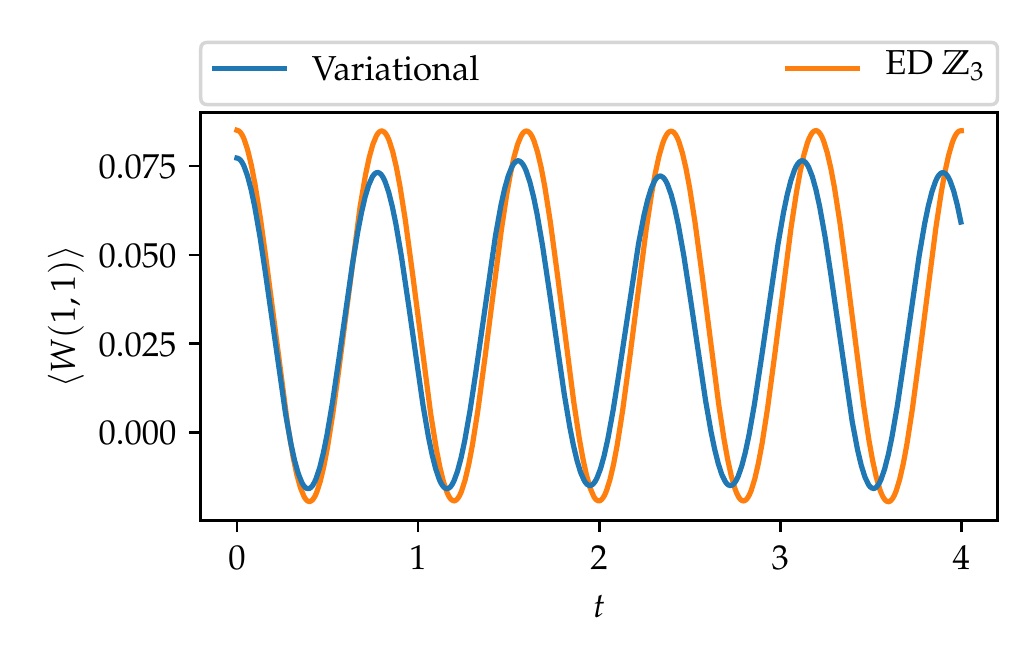}
\caption{Benchmark of the variational time evolution of the $1 \times 1$ Wilson loop after a quench from $g^2=2.5$ to $g^2=4.0$ on a $3 \times 3$ lattice. It is compared with the time evolution of $\mathbb{Z}_3$ lattice gauge theory computed by exact diagonalization (the truncation from $U(1)$ to $\mathbb{Z}_3$ should only play a minor role in the strong-coupling regime).}
\label{benchmark_time}
\end{figure}
Since we are dealing with a variational ansatz, one should try to test it against exact results. 
For a comparison, we use the exact diagonalization results of the $\mathbb{Z}_{3}$ theory. 
Since the truncation in the electric basis led to  significant differences in the ground state energy already for intermediate coupling and time-dynamics increase the variance in the electric field, we can only expect reasonable agreement for a quench within the strong coupling region. 
We choose to quench the Hamiltonian from $g^2=2.5$ to $g^2=4.0$. The result is shown in Fig.~\ref{benchmark_time}. Even though truncation effects might still play a minor role in that quench, the comparison shows that the variational state can approximate amplitude and frequency of the oscillation.

\subsection{Quench dynamics}
\begin{figure}
\centering
\includegraphics[width=\columnwidth]{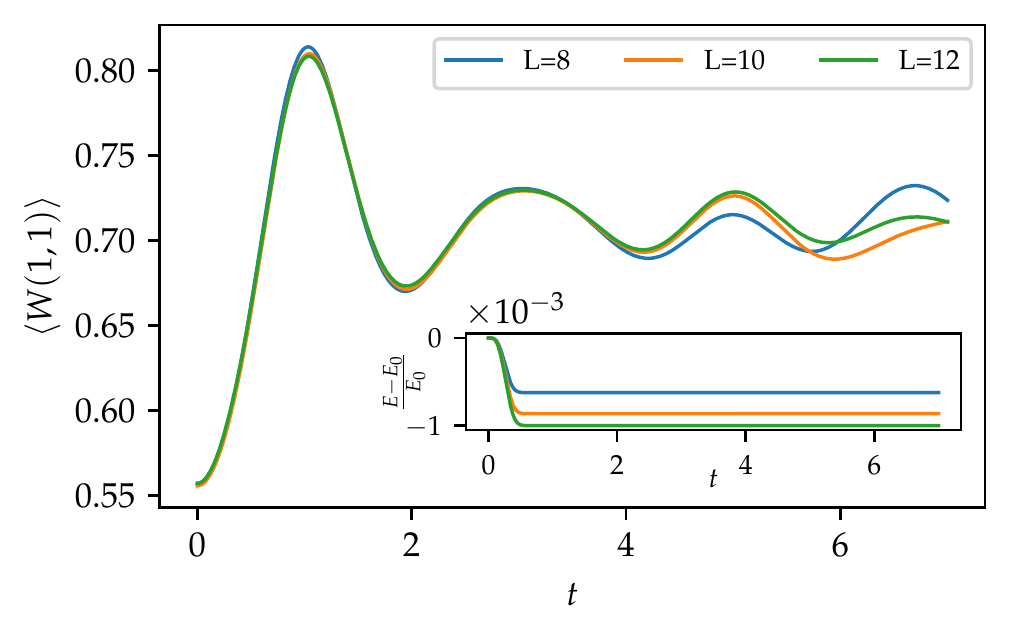}
\caption{Variational time evolution after a quench from $g^2=0.8$ to $g^2=0.5$ for lattice sizes of $8 \times 8$, $10 \times 10$ and $12 \times 12$. The inset shows the relative error in energy $E$ with respect to the initial energy $E_0$ after the quench.} 
\label{smallquench81012}
\end{figure}

\begin{figure}
\centering
\includegraphics[width=\columnwidth]{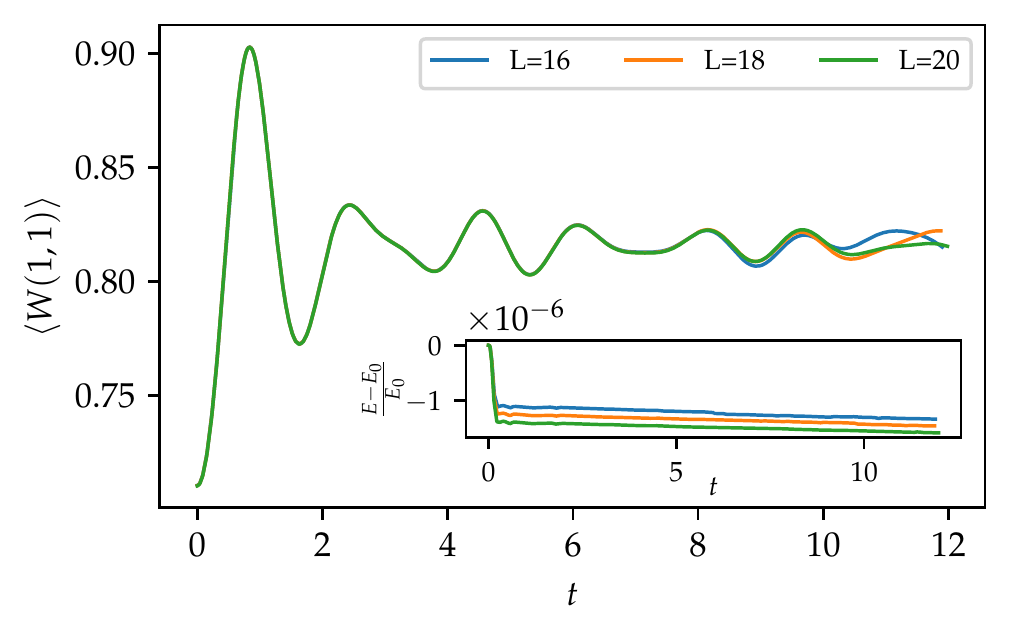}
\caption{Variational time evolution after a quench from $g^2=0.6$ to $g^2=0.3$ for lattice sizes of $16 \times 16$, $18 \times 18$ and $20 \times 20$. The inset shows the relative error in energy $E$ with respect to the initial energy $E_0$ after the quench.}
\label{smallquench161820}
\end{figure}
We start with quenches in the weak-coupling regime where finite-size effects are most pronounced. We are interested in the maximal time up to which we can extract physics in the thermodynamic limit before boundary effects due to our finite lattice start to play a role. 
To compute that point in time, we perform the same quench on different lattice sizes and check where they start to deviate from each other. 
In order to easily compare observables for different lattice sizes, we restrict ourselves to the sector without static charges. We will focus on tracking the $1 \times 1$ Wilson loop during time evolution. 
We probed two different quenches, one from $g^2=0.8$ to $g^2=0.5$ for an $8 \times 8$, $10 \times 10$ and $12 \times 12$ lattice (shown in Fig.~\ref{smallquench81012}) and another one from $g^2=0.6$ to $g^2=0.3$ for lattice sizes of $16 \times 16$, $18 \times 18$ and $20 \times 20$ (shown in Fig.~\ref{smallquench161820}). 
The time evolution on the $8 \times 8$ lattice agrees with the $12 \times 12$ lattice up to $t_{\mathrm{max},8} \sim 3.8$, the $10 \times 10$ lattice up to $t_{\mathrm{max},10} \sim 4.8$. The energy is conserved for all lattice sizes up to a relative error of the order $10^{-3}$. During the time spans where we can reliably extract the time evolution, the Wilson loops indicate equilibrating behavior. This statement is supported by the second quench, where the smaller coupling constants allow us to reach larger lattices. The $16 \times 16$ and $18 \times 18$ lattice agree with the $20 \times 20$ lattice up to $t_{\mathrm{max},16} \sim 8.5$ and $t_{\mathrm{max},18} \sim 9.5$. The energy is conserved up to a relative error of $10^{-6}$. We can only make a statement about the equilibration of Wilson loops since we do not have access to thermal expectation values. 
An interesting direction for future research would be to check whether the Wilson loops thermalize. 
For the calculation of thermal expectation values one could use Monte-Carlo simulations which have been proven successful in computing thermal properties in lattice gauge theory \cite{coddington1986deconfining,chernodub2001lattice}.

\begin{figure}
	\centering
	\includegraphics[width=\columnwidth]{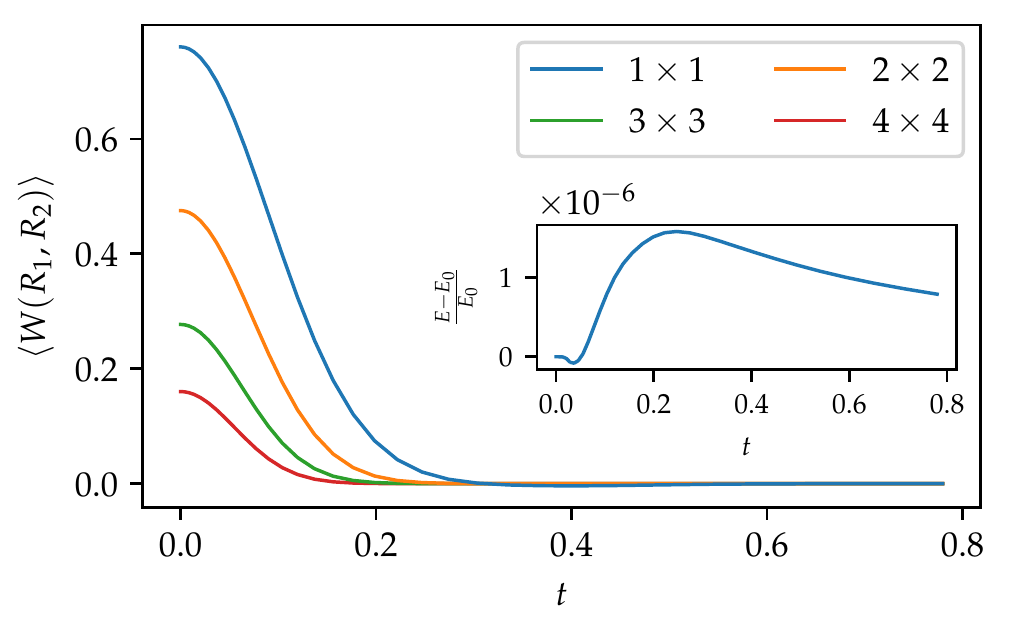}
	\caption{Variational time evolution of the $1 \times 1$, $2 \times 2$, $3 \times 3$ and $4 \times 4$ Wilson loop after a quench from $g^2=0.5$ to $g^2=4.0$ on an $8 \times 8$ lattice. The inset shows the relative error in energy $E$ with respect to the initial energy $E_0$ after the quench.}
	\label{weaktostrong}
\end{figure} 
In the next step, we look at a quench from weak to strong coupling ($g^2=0.5$ to $g^2=4.0$) for an $8 \times 8$ lattice without static charges. 
We track the time evolution of quadratic Wilson loops with edge sizes ranging from one to four. The result is shown in Fig.~\ref{weaktostrong}. 
Although all Wilson loops equilibrate at zero on short time scales (between  $t_{\mathrm{eq},4} \sim 0.2$ for the $4 \times 4$ Wilson loop and $t_{\mathrm{eq},1} \sim 0.5$ for the $1 \times 1$ Wilson loop), we carried out the same evolution on a $7 \times 7$ lattice and found the same behavior. 

The coupling constant at $g^2=4.0$ is large enough to approximate the spectrum by the strong-coupling limit $g^2 \to \infty$, where the eigenstates $\ket{n}$ become diagonal in the electric basis (this can be seen e.g. in the spectrum of the $\mathbb{Z}_3$ theory which is available due to exact diagonalization). In this limit the thermal expectation value of Wilson loops vanishes trivially:

\begin{align}
\expval{W(C)}_{\text{th}}= \frac{1}{Z} \sum_{n} e^{-\beta E_n} \expval**{\prod_{\mathbf{p} \in C}  \frac{1}{2}(U_{\mathbf{p}} + U_{\mathbf{p}}^{\dagger} ) }{ n} = 0.
\end{align}  
For this special quench, we can thus verify that the Wilson loops equilibrate at their thermal expectation value. 

\begin{figure}
	\centering
	\includegraphics[width=\columnwidth]{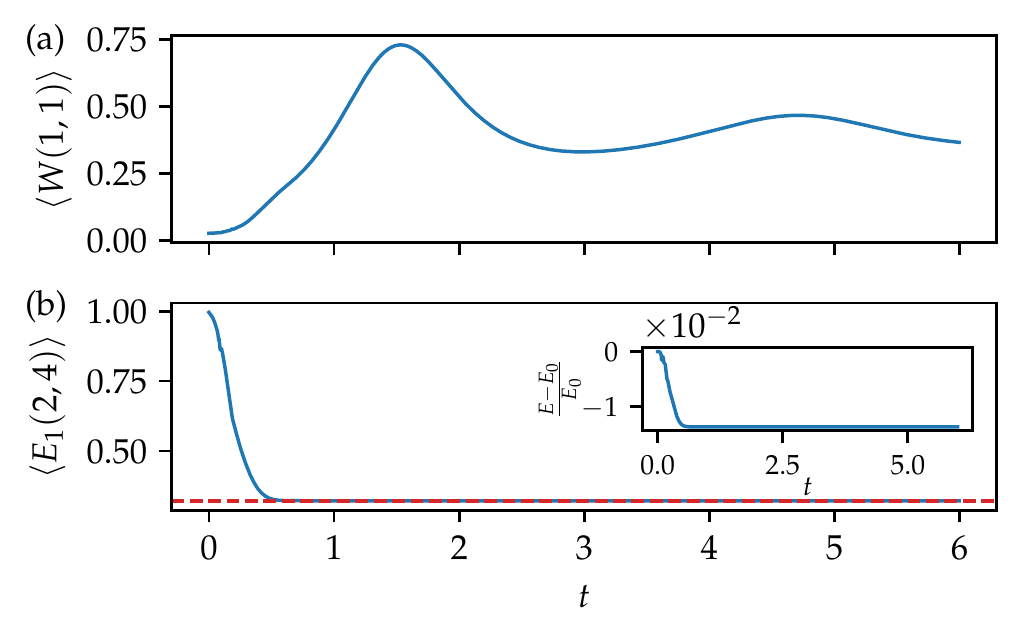}
	\caption{Variational time evolution on an $8 \times 8$ lattice after a quench from $g^2=4.0$ to $g^2=0.5$ with a positive charge placed at ($x_1=2,x_2=4$) and a negative charge at ($x_1=6,x_2=4$). We measure (a) the $1 \times 1$ Wilson loop at the origin $W(1,1)$ and (b) the electric field on a link between the two charges $E_1(2,4)$. The red dashed line represents the Coulomb value of the electric field. The inset shows the relative error in energy $E$ with respect to the initial energy $E_0$ after the quench.}
	\label{strongtoweak}
\end{figure}

The next quench we will study is from strong to weak coupling. We quench on an $8 \times 8$ lattice from $g^2=4.0$ to $g^2=0.5$ with static charges horizontally separated by four links. 
Besides the $1 \times 1$ Wilson loop at the origin, we observe how the electric field of the ground state at $g^2=4.0$, a strongly confined fluxtube, evolves after the quench, in particular the electric field $E_1(x_1=2,x_2=4)$ (one of the links inside the fluxtube, see Fig.~\ref{strongtoweak}). It starts close to one, the strong-coupling value of the electric field, and decreases rapidly to $E_1^C(2,4)=0.322$, the value of the Coulomb electric field on that link (shown in the red dashed line). 
The Wilson loop seems to equilibrate on longer time scales. 

The energy is conserved up to a relative error of $10^{-2}$. 
The larger error compared to previous quenches can be explained by the fact that around $t \sim 0.25$ the approximation method of the infinite sums appearing in the evaluation of expectation values changes from the low $\gamma_{\mathbf{k}}$ to the high $\gamma_{\mathbf{k}}$ approximation (see section~\ref{variationalansatz}). 
In that transition region higher orders need to be calculated using uniform sampling (see Appendix~\ref{evaluation}) which introduces additional errors. 
However, the relative error is still small and observables have no visible jump in this region, indicating that the two approximation schemes work. 
After the transition region the energy is well conserved due to the fact that the variational parameters $\gamma_{\mathbf{k}}^{I}$ increase, making the approximation of the infinite sums involved in the calculation of expectation values very easy (see section~\ref{variationalansatz}). 

\begin{figure*}
	\centering
	\includegraphics[width=\textwidth]{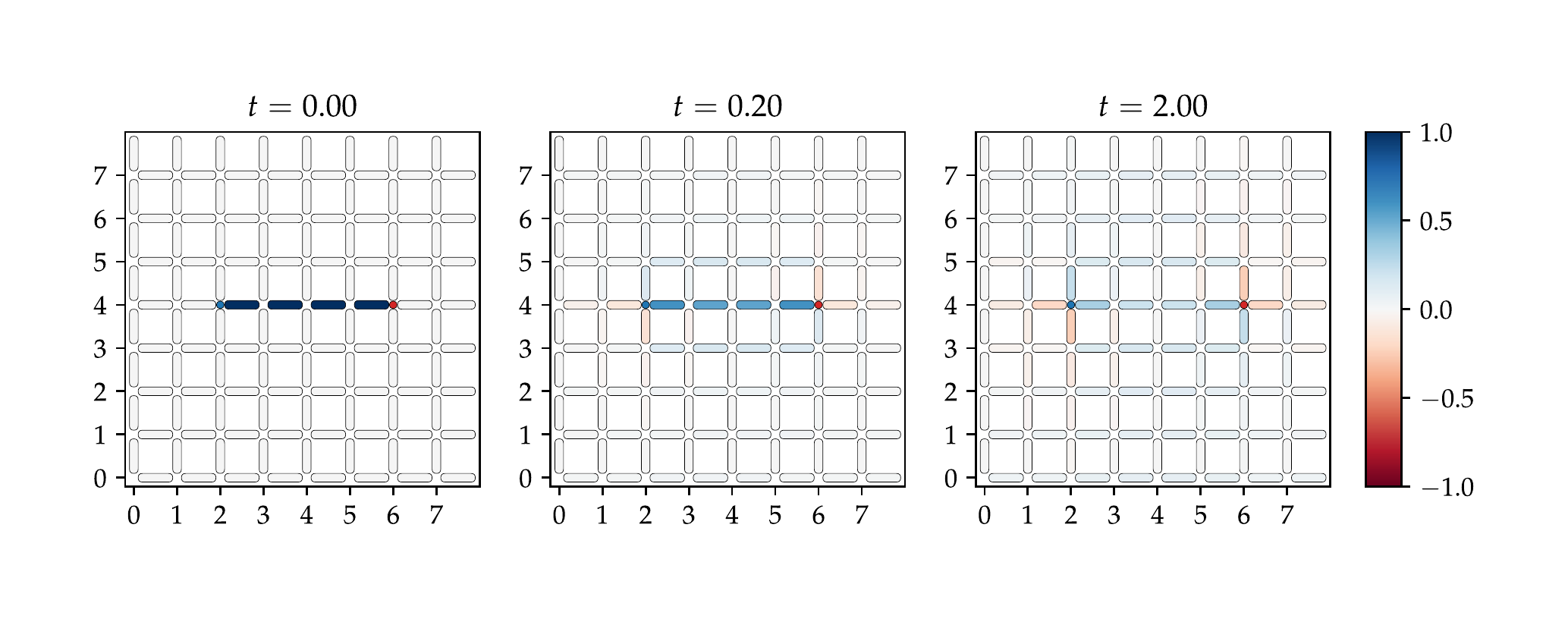}
	\caption{Variational time evolution of the electric field on an $8 \times 8$ lattice after a quench from $g^2=4.0$ to $g^2=0.5$ with a positive charge placed at ($x_1=2,x_2=4$) (blue dot) and a negative charge at ($x_1=6,x_2=4$) (red dot). The color of the charges is only for graphical illustration (not related to the colorbar). The expectation value of the electric field is shown at $t=0.0$, $t=0.2$ and $t=2.0$. At $t=0.0$, the state is in the variational ground state for $g^2=4.0$ where the electric flux is confined between the two charges. After the quench, the electric field starts to spread over the lattice ($t=0.2$) and equilibrates at the Coulomb value for this charge configuration ($t=2.0$).}
	\label{strongtoweakefield}
\end{figure*}

The spreading of the electric field from inside the flux tube between the two charges towards the Coulomb configuration of the electric field is illustrated in Fig.~\ref{strongtoweakefield}. An interesting question is whether the state becomes deconfined at long times. We cannot use the scaling of spatial Wilson, this only serves as an indicator for confinement in the ground state \cite{svetitsky1982critical}. Since in our formulation the value of the longitudinal (Coulomb) part of the electric field is fixed and only the transversal part is dynamical (see Appendix~\ref{appformulation}), we can measure precisely how much an electric field configuration differs from the Coulomb configuration. At $t=2.0$, in the last of the three pictures in Fig.~\ref{strongtoweakefield}, the difference to the Coulomb configuration is of order $10^{-12}$ for the whole lattice, with no remnant of an electric flux tube between the two charges. This is a strong indication that the state becomes deconfined, corresponding possibly to a thermal state with a temperature above the confinement-deconfinement transiton~\cite{parga1981finite,svetitsky1986symmetry}.

\section{\label{conclusion} Conclusion}
We introduce a new class of variational states, complex periodic Gaussian states, to study ground state properties and real-time dynamics in a (2+1)-dimensional $U(1)$ lattice gauge theory. 
The evaluation of expectation values can only partially be done analytically, an infinite sum remains to be computed numerically. 
We present a scheme to approximate them for all variational parameters on an $8 \times 8$ lattice and for the weak-coupling regime up to $20 \times 20$. 
This allows us to study the variational ground state of these states over the whole coupling region and extract the thermodynamic limit. We benchmark our ansatz against the exact ground state for the one-plaquette case.
We also compute the string tension using two different methods: First, by fitting the static potential between two charges with a 2d Coulomb potential and a linear potential. Secondly, we fit the exponential decay of Wilson loops with an area and a perimeter law. 
The two approaches are complementary since in the strong-coupling regime Wilson loops become difficult to fit due to the tiny value of large Wilson loops while the static potential approach works well as energy differences become larger. In the weak-coupling regime, however, the string tension becomes too small to extract the linear part of the potential on the given lattice sizes while Wilson loops decay only modestly allowing reliable fits. We are able to observe the expected exponential decay of the string tension in the weak-coupling regime.

Since our variational states do not need a truncation in the local Hilbert space, we compare our $U(1)$ ground state data with exact diagonalization results for a $\mathbb{Z}_{3}$ theory to study trunctation effects in the electric basis. 
The results agree for strong couplings and start to differ significantly for intermediate couplings. While the ground state energy of the truncated theory goes to zero in the continuum limit $g^2 \to 0$ (since the electric energy is bounded), the variational ground state energy tends towards a finite value due to the variance of the electric field growing unboundedly.

In the last section, using the time-dependent variational principle, we probe out-of-equilibrium dynamics after a quench of the coupling constant. As a benchmark, we compare the variational time evolution after a quench within the strong-coupling regime with exact diagonalization results of the $\mathbb{Z}_3$ theory. We then start by studying quenches within the weak-coupling regime where we expect finite size effects to be significant. We compare the time evolution of a Wilson loop after the same quench for different lattice sizes to estimate at which time scales smaller lattices deviate from the thermodynamic limit. The times we can reach are large enough to indicate equilibration of Wilson loops. 

In the next step, we perform a quench from weak ($g^2=0.5)$ to strong coupling ($g^2=4.0$) and track the time evolution of differently sized Wilson loops. They all equilibrate at zero which is the thermal expectation value in the strong-coupling limit ($g^2 \to \infty$). Since the spectrum at $g^2=4.0$ is close to the strong-coupling limit, this indicates that the Wilson loops equilibrate at their thermal expectation values. 
We also study a quench from strong to weak coupling in the sector of two static charges. We observe that the electric flux, which is perfectly confined for the strong-coupling ground state, spreads over the whole lattice and equilibrates at the Coulomb value for the electric field to very high accuracy, leaving no trace of confinement. 

In all considered quenches, we see equilibrating behavior of observables up to the times where boundary effects start to play a role. It would be interesting to compare the equilibrated expectation values to thermal expectation values which can be computed by Monte-Carlo simulations \cite{coddington1986deconfining,chernodub2001lattice}. Another interesting application for Monte-Carlo methods would be in the numerical evaluation of the variational ansatz by approximating the infinite sums. This could potentially enable the simulation of larger system sizes. The accuracy of these simulations would need to be high in order to carry out the evolution over reasonable time scales while ensuring energy conservation. Another natural extension of this work is the treatment of (3+1)-dimensional compact QED. By generalizing an idea in Ref.~\cite{drell1979quantum} to complex Gaussian states, a variational ansatz can be designed for 3+1 dimensions. However, due to additional local constraints appearing in 3+1 dimensions (compared to one global constraint in 2+1 dimensions), a new numerical approximation scheme would be required. Another interesting idea is to include dynamical matter.To couple the gauge degrees of freedom to matter, it is essential to ﬁnd a formulation of such a theory, which admits the same gauge-invariant variables as used in this work for static matter. Recently, such a formulation has been proposed \cite{bender2020gauge}. This could allow to combine a periodic Gaussian state for the gauge ﬁeld with a fermionic ansatz state, describing dynamical matter.
Extending the ansatz to non-Abelian gauge theories is more difficult since they do not allow a translationally invariant formulation in terms of gauge-invariant plaquette variables. However, other gauge-invariant variables could be used to construct similar ansatz states \cite{ligterink2000toward}. 

\acknowledgements
We thank Lorenzo Piroli and Lucas Hackl for helpful discussions. J.B., P.E. and I.C. acknowledge support by the EU-QUANTERA project QTFLAG (BMBF grant No. 13N14780). P.E. acknowledges support from the International Max-Planck Research School for Quantum Science and Technology (IMPRS-QST). J.B. and P.E. thank the Hebrew University of Jerusalem for the hospitality during their stay at the Racah Institute of Physics.

\bibliography{bibliography}

\appendix
\section{\label{appformulation}Formulation in terms of plaquette variables}
In this section, we want to give a short review on the separation of gauge fields into (almost) gauge invariant plaquette variables and a static part corresponding to the longitudinal Coulomb field. 
For simplicity and since this is the charge configuration used throughout the paper, we will focus on a situation with two static charges placed vertically at $x_2=d_2$ separated horizontally by a distance $d$.
Other charge configurations follow analogously. 
We want to split the electric flux line between the two charges into a transversal component, generated by the lattice curl of a field $\epsilon$ on the plaquettes and into a longitudinal component, generated by the lattice gradient of a scalar field $\phi$ on the vertices. 
All other electric flux configurations can be created on top of it by exciting an electric flux loop around a plaquette or around the whole lattice. 

The longitudinal part is by definition of the form

\begin{align} \label{gradient_nabla}
E_{i}^L(\mathbf{x})=  -\nabla^{(+)} \phi(\mathbf{x}) \equiv -(\phi(\mathbf{x} + e_i) - \phi(\mathbf{x}))
\end{align}
where $\nabla^{(+)}$ is the lattice forward derivative. Using Gauss law, 
\begin{align}
\sum_i \nabla^{(-)}_i E_i^L (\mathbf{x})= \sum_i E_i^L(\mathbf{x}) - E_i^L(\mathbf{x}-\mathbf{e}_i) =Q(\mathbf{x})
\end{align}
with $\nabla^{(-)}$ the lattice backward derivative, we arrive at a lattice version of Poisson's equation: 

\begin{align}
- \nabla^{(-)}\nabla^{(+)} \phi (\mathbf{x}) = Q(\mathbf{x})
\end{align}
The solution for $\phi$ is 

\begin{align}
\phi(\mathbf{x}) = \frac{1}{L} \sum_{\mathbf{y}} Q(\mathbf{y}) \sum_{\mathbf{k} \neq \mathbf{0}} \frac{e^{2\pi i \frac{k_1 (x_1-y_1) +k_2(x_2-y_2)}{L}}}{4 - 2 \cos\left(\frac{2\pi k_1}{L}\right) - 2 \cos\left(\frac{2\pi k_2}{L}\right)}
\end{align}
with $\mathbf{x}=(x_1,x_2)$ and $x_1,x_2$ ranging from $0$ to $L-1$. 
The same applies to $\mathbf{y}$ and $\mathbf{k}$. 
There is no $\mathbf{k}=\mathbf{0}$ contribution since the total charge on a periodic lattice needs to be zero because of gauge invariance. 
$E_i^L(\mathbf{x})$ then follows straightforwardly from~\eqref{gradient_nabla}.

We write the transversal part as the curl of an $\epsilon$-field on the plaquettes, 

\begin{align}
E_i^T (\mathbf{x}) = \nabla^{(-)} \times \epsilon \equiv \epsilon_{ij} \nabla^{(-)}_j \epsilon(\mathbf{x}) 
\end{align}
where the plaquette corresponding to $\mathbf{x}$ is the one having $\mathbf{x}$ as its the bottom left corner.
We take the curl of the above expression and use a lattice analog of the vector identity $\nabla \times \nabla \times A = \nabla (\nabla \cdot A) - \Delta A$, here in two dimensions, to obtain a Poisson equation for the $\epsilon$-field:

\begin{align}
-\nabla^{(+)}\nabla^{(-)} \epsilon (\mathbf{x}) = \epsilon_{ij} \nabla_{i}^{(+)} E_{j}(\mathbf{x})
\end{align}
where we sum over repeated indices. 
This equation is solved by  
\begin{equation}
\begin{aligned}
    \epsilon(\mathbf{x}) = &\frac{1}{L} \sum_{\mathbf{y}} \epsilon_{ij} \nabla_{i}^{(+)} E_{j}(\mathbf{y}) \times\\
    &\times\sum_{\mathbf{k} \neq \mathbf{0}} \frac{e^{2\pi i \frac{k_1 (x_1-y_1) +k_2(x_2-y_2)}{L}}}{4 - 2 \cos\left(\frac{2\pi k_1}{L}\right) - 2 \cos\left(\frac{2\pi k_2}{L}\right)}
\end{aligned}
\end{equation} 
In our case, $\epsilon_{ij} \nabla_{i}^{(+)} E_{j}(\mathbf{x})$ is $1$ on the plaquettes above the electric string connecting the charges and $-1$ below the string. 
However, since the sum over this expression will always be zero, we cannot generate a constant electric field with the $\epsilon$-field which is required since $\tilde{E}_1(\mathbf{k=0})=\frac{d}{L}$. 
It is important to also consider the Polyakov loop winding horizontally around the lattice. 
We choose it to wind around the lattice at $x_2=d_2$ and the electric field along it to be $\epsilon_{\mathrm{poly},1}=\frac{d}{L}$.  
We define an additional $\epsilon$-field $\epsilon_{\mathrm{const}}$ on the plaquettes, on top of $\epsilon$: 

\begin{align}
\epsilon_{\mathrm{const}} (\mathbf{x})=
\begin{cases} 
\frac{d}{L^2} (x_2-d_2) & x_2 \geq d_2 \\
\frac{d}{L} - (d_2 - x_2) \frac{d}{L^2} & x_2 < d_2 \\ 
\end{cases}
\end{align}
It is defined in such a way that 

\begin{align}
E_{\mathrm{const},1}(\mathbf{x})=\nabla^{(-)} \times \epsilon_{\mathrm{const}} + \epsilon_{\mathrm{poly},1}  \delta_{x_2,d_2} = \frac{d}{L^2}
\end{align}
giving us the $\mathbf{k}=0$ component of the electric field. 
We can now rewrite the electric field operator as:

\begin{align}
\hat{E}_{i}(\mathbf{x})=&\big(\nabla^{(-)} \times (\hat{L}(\mathbf{x})+\epsilon(\mathbf{x})+\epsilon_{\mathrm{const}}(\mathbf{x}))\big)_i \nonumber\\
&+ \delta_{i,1} \delta_{x_2,d_2} (\hat{L}_{\mathrm{poly},1}+\epsilon_{\mathrm{poly},1})+ \delta_{i,2} \delta_{x_1,d_1} \hat{L}_{\mathrm{poly},2} \nonumber\\
&+ E_{i}^{L}(\mathbf{x}).
\end{align}
$d_1$ is the $x_1$-position where the Polyakov loop winds vertically around the lattice. 
The operators $\hat{L}(\mathbf{x})$ and $\hat{L}_{\mathrm{poly}}$ measure the electric flux around a plaquette, resp. around the lattice, on top of the contributions given by the charge configuration. 
Their eigenvalues are integer-valued. 
If we insert the restriction to the topological sector with $L_{\mathrm{poly},1}=L_{\mathrm{poly},2}=0$ and define $E^C_i(\mathbf{x})=E^L_i(\mathbf{x}) + \delta_{i,1} \frac{d}{L^2}$ as the Coulomb electric field, we obtain the electric Hamiltonian given in eq.~\eqref{Hamiltonianref}:

\begin{align}
&\begin{alignedat}{2}
H_{E}&=\frac{g^2}{2} \sum_{\mathbf{x},i} \big(E^C_i(\mathbf{x}) + \epsilon_{ij} (&&L(\mathbf{x})-L(\mathbf{x}-\mathbf{e}_j) \nonumber\\
&&&+\epsilon(\mathbf{x}) - \epsilon(\mathbf{x}-\mathbf{e}_{j})) \big)^2 \nonumber
\end{alignedat}\\
&\phantom{H_E}=E_C +\frac{g^2}{2} \sum_{\mathbf{x},i} (L(\mathbf{x})-L(\mathbf{x}-\mathbf{e}_i) +\epsilon(\mathbf{x}) - \epsilon({\mathbf{x}-\mathbf{e}_{i}}) )^2
\end{align}
with the Coulomb energy $E_C=\frac{g^2}{2} \left(\frac{d^2}{L^2} + \sum_{\mathbf{x},i} E_i^L(\mathbf{x}) \right)$. 
Besides orthogonality between longitudinal and transversal component of the electric field, we also used Plancherel's theorem which ensures orthogonality between the constant part and the other two since their $\mathbf{k}=0$ component is zero.

\section{\label{evaluation} Numerical evaluation of complex periodic Gaussian states} \label{evaluation ansatz}
In this section, we review the numerical evaluation of complex periodic Gaussian states in more detail. 
We saw in section~\ref{variationalansatz} that the region with $\gamma_{\mathbf{k}}=\gamma_{\mathbf{k}}^R + \left(\gamma_{\mathbf{k}}^I\right)^2\left(\gamma_{\mathbf{k}}^R\right)^{-1} \approx 1$ is the most difficult to evaluate. 
Since the variational ground state (for which $\gamma_{\mathbf{k}}^I = 0$) varies from high $\gamma_{\mathbf{k}}^R$ for low couplings to low $\gamma_{\mathbf{k}}^R$ for large couplings, there is a transition region at $g^2 \sim 1.1$ where $\gamma_{\mathbf{k}}$ approaches one. 
We therefore want to study the approximations to all infinite sums involved in the computation of the variational energy in eq.~\eqref{varenergy} on an $8 \times 8$ lattice without static charges for the ground state at $g^2=1.1$ which is the highest coupling where the high $\gamma_{\mathbf{k}}$ approximation is used and $g^2=1.2$ which is the lowest coupling for which the low $\gamma_{\mathbf{k}}$ approximation is used. 
For all other couplings the contributions to infinite sums decay faster with higher orders compared to one of the two examples discussed below. 
The variational parameters $\gamma_{\mathbf{k}}^R$ for these states (rounded to three digits) are shown in Table~\ref{varparam1} and~\ref{varparam2}. 
The values are displayed not only for the independent $\gamma_{\mathbf{k}}^R$ ($\mathbf{k} \in \mathcal{K}$), but are split between the dependent parameters $\gamma_{\mathbf{k}}^R$ and $\gamma_{\mathbf{-k}}^R$, to illustrate that they lie in the transition region $\gamma_{\mathbf{k}} \approx 1$ between the two approximation methods. \\
\begin{table} 
	\caption{Variational parameters $\gamma_{\mathbf{k}}^R$ for the variational ground state at $g^2=1.1$ } \label{varparam1}
	\begin{tabular}{c|c|c|c|c|c|c|c|c} 
		\diagbox{$k_x$}{$k_y$}   &    0     & 1   &   2   &   3  &    4   &   5    &  6  &    7 \\
		\hline
		0  &       &1.289  &1.108  &0.978  &0.937  &0.978  &1.108  &1.289\\
		1  &1.288  &1.206  &1.050  &0.937  &0.897  &0.936  &1.051  &1.207\\
		2  &1.109  &1.050  &0.937  &0.849  &0.819  &0.849  &0.936  &1.050\\
		3  &0.979  &0.935  &0.849  &0.781  &0.756  &0.781  &0.849  &0.937\\
		4  &0.935  &0.897  &0.819  &0.756  &0.732  &0.756  &0.819  &0.897\\  
		5  &0.979  &0.937  &0.849  &0.781  &0.756  &0.781  &0.849  &0.935\\
		6  &1.109  &1.050  &0.936  &0.849  &0.819  &0.849  &0.937  &1.050\\
		7  &1.288  &1.207  &1.051  &0.936  &0.897  &0.937  &1.050  &1.206\\
	\end{tabular}

	\bigskip
	\caption{Variational parameters $\gamma_{\mathbf{k}}^R$ for the variational ground state at $g^2=1.2$  } \label{varparam2}
	\begin{tabular}{c|c|c|c|c|c|c|c|c} 
		\diagbox{$k_x$}{$k_y$}   &    0  &    1  &    2  &    3  &    4  &    5  &    6  &    7 \\
		\hline
		0  &       &1.140  &1.002  &0.889  &0.852  &0.889  &1.002  &1.140\\ 
		1  &1.137  &1.078  &0.944  &0.855  &0.826  &0.855  &0.949  &1.075\\
		2  &0.999  &0.956  &0.858  &0.779  &0.753  &0.783  &0.859  &0.946\\
		3  &0.893  &0.853  &0.783  &0.721  &0.699  &0.721  &0.785  &0.857\\
		4  &0.862  &0.823  &0.752  &0.695  &0.676  &0.695  &0.752  &0.823\\
		5  &0.893  &0.857  &0.785  &0.721  &0.699  &0.721  &0.783  &0.853\\
		6  &0.999  &0.946  &0.859  &0.783  &0.753  &0.779  &0.858  &0.956\\
		7  &1.137  &1.075  &0.949  &0.855  &0.826  &0.855  &0.944  &1.078\\
	\end{tabular} 
\end{table}

\begin{figure*}[t]
\centering
\subfloat[][]{
	\includegraphics[width=\columnwidth]{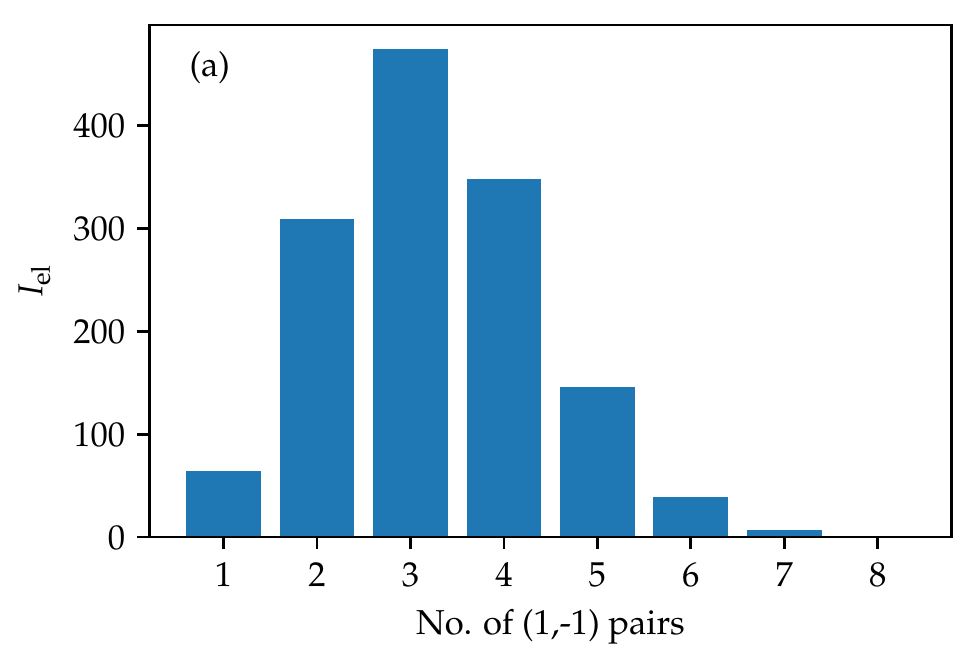}
}
\hfill
\subfloat[]{
	\includegraphics[width=\columnwidth]{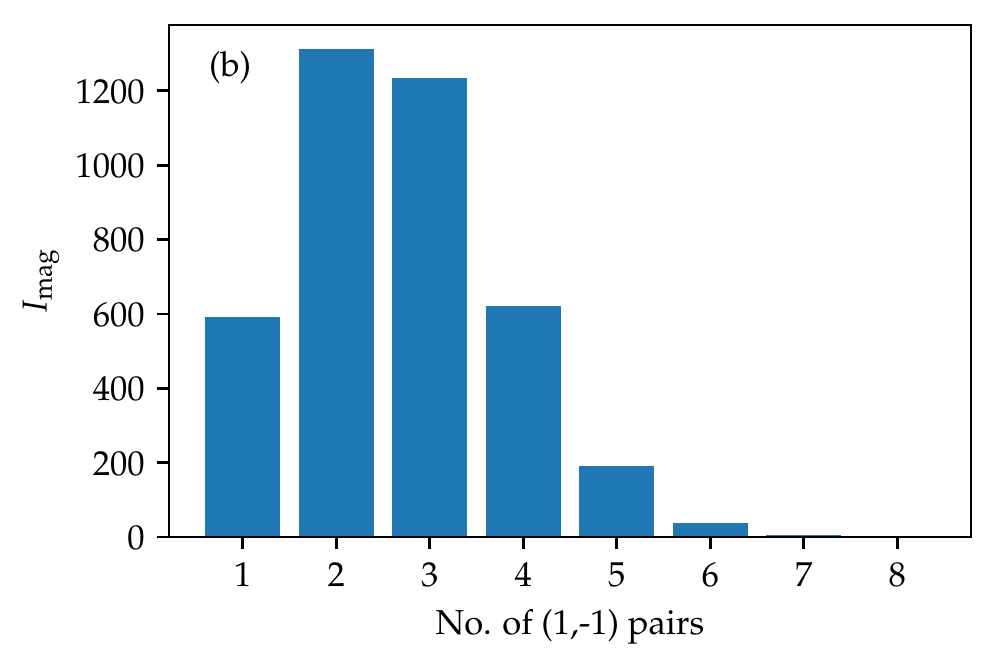}
}
	\caption{Contributions of different orders of $N_{\mathbf{p}}$ configurations to the infinite sums $I_{el}$ (a) and $I_{mag}$ (b) appearing in the high $\gamma_{\mathbf{k}}$ approximation of the variational energy. Due to the constraint $N_{\mathbf{k}=0}=0$, the sum over all elements of $N_{\mathbf{p}}$ needs to be zero. Every bar represents the summed contributions of all $N_{\mathbf{p}}$ configurations containing a certain number of (1,-1) pairs and the remaining entries zero. Orders which are not of this type have a negligible contribution, e.g. $\{N\}_{-2,1,1}$  has a summed contribution to $I_{el}$ of $0.076$ and a summed contribution to $I_{mag}$ of $0.23$. }
	\label{order_truncation}
\end{figure*}
\begin{figure*}[t]
	\centering
    \subfloat[][]{
    	\includegraphics[width=\columnwidth]{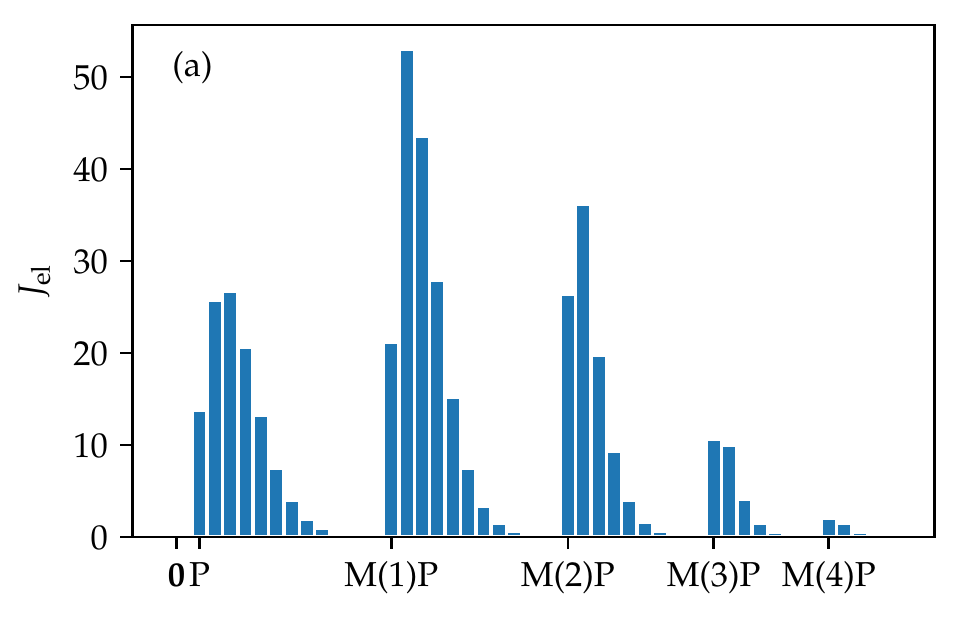}
    }
    \hfill
    \subfloat[]{
    	\includegraphics[width=\columnwidth]{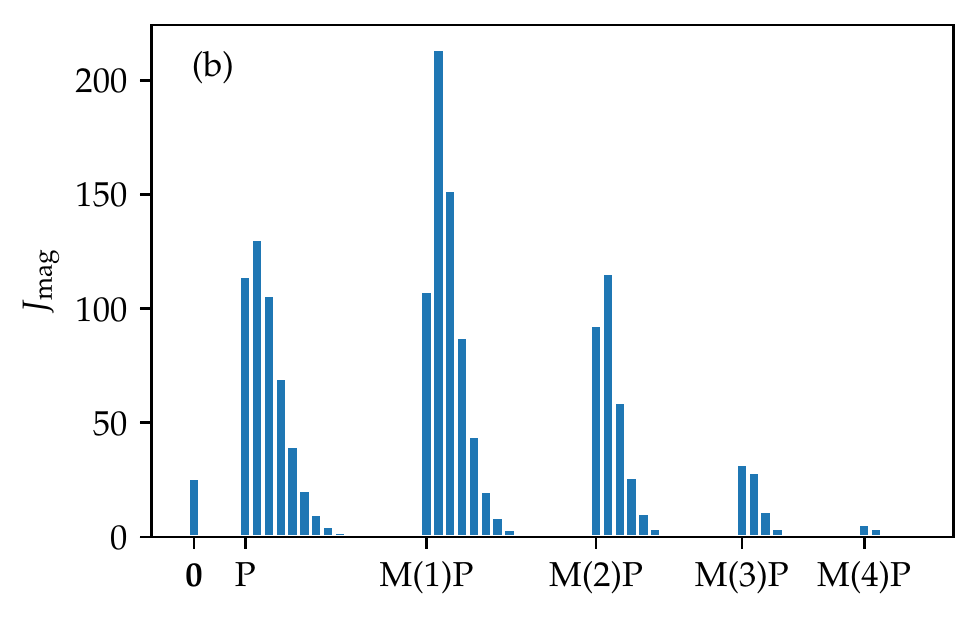}
    }
	\caption{Contributions of different orders of $N_{\mathbf{p}}$ configurations to the infinite sums  $J_{el}$ (a) and $J_{mag}$ (b) appearing in the low $\gamma_{\mathbf{k}}$ approximation of the variational energy. Due to absence of a constraint, all $N_{\mathbf{p}}$ configurations need to be considered. The orders are organized in groups. $P$ denotes orders which contain a growing number of $1$'s. Since $N_{\mathbf{p}}$ and $-N_{\mathbf{p}}$ are evaluated together, $P$ also represents orders with a growing number of $-1$'s. $M(1)P$ contains orders whose non-zero elements are a single $-1$ and a growing number of $1$'s. The first order in $M(1)P$ contains a pair of (1,-1) as non-zero elements. $M(2)P$ is structured in the same way as $M(1)P$ but with two $-1$'s. Analogously for the other groups. The $N_{\mathbf{p}}=\mathbf{0}$ configuration (denoted as $\mathbf{0}$) has vanishing contribution to $J_{el}$ but a non-zero contribution to $J_{mag}$.}
	\label{inverse_order_truncation}
\end{figure*}
Using $\gamma_{\mathbf{k}}^I=0$ and $\epsilon_{\mathbf{p}}=0$, the expressions we need to compute for the variational ground state at $g^2=1.1$ simplify significantly: 
\begin{equation}
\begin{aligned}
I_{el} \equiv &\sum_{ \{ N_{\mathbf{k}=\mathbf{0}}=0 \}}   e^{- \pi \sum_{\mathbf{k}} |N_{\mathbf{k}}|^2  \gamma_{\mathbf{k}}^R } \sum_{\mathbf{k}} \left(\gamma_{\mathbf{k}}^R\right)^2 | N_{\mathbf{k}}|^2\times\\
&\times \left(4-2 \cos\left(\frac{2\pi k_{1}}{L}\right)-2 \cos\left(\frac{2\pi k_{2}}{L}\right) \right) 
\end{aligned}
\end{equation}
for the electric energy,
\begin{align}
I_{mag} \equiv \sum_{ \{ N_{\mathbf{k}=\mathbf{0}}=0 \}}  e^{- \pi \sum_{\mathbf{k}} |N_{\mathbf{k}}|^2  \gamma_{\mathbf{k}}^R }  \sum_{\mathbf{p}} (-1)^{N_{\mathbf{p}}} 
\end{align}
for the magnetic energy and the normalization 
\begin{align}
I_0= \sum_{ \{ N_{\mathbf{k}=\mathbf{0}}=0 \}}  e^{- \pi \sum_{\mathbf{k}} |N_{\mathbf{k}}|^2  \gamma_{\mathbf{k}}^R }.
\end{align}
We include orders with $N_{\mathbf{p}}$ configurations of up to 8 pairs of $\{1,-1 \}$ and the rest zeros. 
The first three are computed exactly and the remaining five by uniform sampling. 
Additionally, we compute exactly the orders $\{N \}_{2,-1,-1}$ and $\{N\}_{-2,1,1}$ to show they have negligible contributions. 
Orders like these, whose $N_{\mathbf{p}}$ configurations differ only by a minus sign can be evaluated together by evaluating for every permutation not only the contribution of $N_{\mathbf{p}}$ but also of $-N_{\mathbf{p}}$. 
Therefore, from now on orders which are not closed under reflection will also include all their permutations multiplied by minus one. 
This will be heavily used in the low $\gamma_{\mathbf{k}}$ approximation. 

The exact evaluation of orders is based on an algorithm which generates all permutations of a multiset in $O(1)$ time~\cite{williams2009loopless}, i.e. the time to generate a new permutation is independent of the permutation size. 
It is much smaller than the time needed to do computations with a permutation which allows to highly parallelize the process and reach higher orders. 
The evaluation of an observable with respect to a set of permutations $\{ N \}$ with uniform sampling is based on the approximation: 

\begin{align}
\sum_{N_{\mathbf{p}} \in \{N\}} O(N_{\mathbf{p}}) \approx \frac{p}{s} \sum_{N_{\mathbf{p}} \in S} O({N_{\mathbf{p}}})
\end{align}
where $S$ is a set of $s$ randomly drawn $N_{\mathbf{p}}$ configurations from $\{N \}$ and $p$ the number of permutations within this order. 
For all orders which are computed with uniform sampling we use $s=10^8$ in the high $\gamma_{\mathbf{k}}$ approximation and $s=10^7$ in the low $\gamma_{\mathbf{k}}$ approximation. 
The contributions to $I_{el}$ and $I_{mag}$ for the high $\gamma_{\mathbf{k}}$ approximation are displayed in Fig.~\ref{order_truncation}. 
We do not show this analysis for the normalization since its contributions decay faster than the ones for $I_{el}$ and $I_{mag}$. 
The errors due to uniform sampling are too small to be shown in the plot, the biggest error occurs in the order with four pairs of $\{1,-1\}$ which has a contribution  of $347.54(15)$ to $I_{el}$ and of $622.70(24)$ to $I_{mag}$.

For the variational ground state at $g^2=1.2$, the infinite sums in eq.~\eqref{varenergy} reduce to
\begin{equation}
    \begin{aligned}
        J_{el}=&\sum_{ \{ N_{\mathbf{p}}  \}/\sim_1 } e^{- \pi \sum_{\mathbf{k}} |N_{\mathbf{k}}|^2 \left(\gamma_{\mathbf{k}}^R\right)^{-1} } \times\\
        &\times\sum_{\mathbf{k}} \left(4-2 \cos\left(\frac{2\pi k_{1}}{L}\right)-2 \cos\left(\frac{2\pi k_{2}}{L}\right) \right) |N_{\mathbf{k}}|^2
    \end{aligned}
\end{equation}
for the computation of the electric energy,
\begin{align}
J_{mag}=\sum_{ \{ N_{\mathbf{p}}  \}/\sim_1 } \sum_{\mathbf{p}} e^{-\pi \sum_{\mathbf{k}} |N_\mathbf{k} - \frac{1}{2}^{\mathbf{p}}_\mathbf{k}|^2  \left(\gamma_{\mathbf{k}}^R\right)^{-1} } 
\end{align}
with $\frac{1}{2}^{\mathbf{p}}_\mathbf{k}= \frac{1}{2L} e^{-i\mathbf{p}\mathbf{k}}$ for the computation of the magnetic energy and
\begin{align}
J_{0}=\sum_{ \{ N_{\mathbf{p}}  \}/\sim_1 } e^{- \pi \sum_{\mathbf{k}} |N_{\mathbf{k}}|^2 \left(\gamma_{\mathbf{k}}^R\right)^{-1} }
\end{align}
for the normalization. 
Since we do not have a global constraint in the low $\gamma_{\mathbf{k}}$ approximation, more orders contribute to the infinite sums. 
The contributions to $J_{el}$ and $J_{mag}$ of different orders are given in Fig.~\ref{inverse_order_truncation}.
The errors are again too small to be displayed, the biggest one occurs in the order $\{N\}_{-1,1,1,1,1,1}$ with contributions of $15.22(1)$ to $J_{el}$ and of $44.07(4)$ to $J_{mag}$. 

Both approximation schemes decay reasonably well with higher orders and the truncation of even higher orders can be justified. 
Moreover, the errors introduced due to uniform sampling are small, in particular since the lowest orders were still calculated exactly. 
The algorithm we applied during computations to decide with which approximation method an expectation value should be evaluated was to select higher orders and compute them by uniform sampling with a low sample size of $s=10^5$. 
This allowed us to choose the scheme which had a better decay with higher orders.

\vspace{2em}
\section{Observables} \label{observables}
In this section, we provide formulas for important quantities which are too lengthy to fit into the main body of the manuscript. 
This includes formulas for expectation values of observables, namely Wilson loops and electric field, and a formula for the gradient of the energy with respect to the variational parameters which is essential to minimize the variational energy and carry out the time-dependent varational principle. 
For the latter, we present additionally a formula for the Gram matrix. For every infinite sum appearing in the expressions, we provide both the high and low $\gamma_{\mathbf{k}}$ approximation.

We start with the expectation value of a Wilson loop along a contour $C$, where $\mathbf{p} \in C$ denotes all plaquettes within this contour:
\begin{widetext}
\begin{align}
\frac{\expval{W(C)}{\Psi_{CPG}}}{\braket{\Psi_{CPG}}}&= e^{-\frac{\pi}{4L^2} \sum_{\mathbf{k \neq 0} } \left(\gamma_{\mathbf{k}}^R\right)^{-1} \sum_{\mathbf{p}, \mathbf{p'}} \cos(\mathbf{k}  (\mathbf{p}-\mathbf{p'})) }  \expval{\prod_{\mathbf{p} \in C}(-1)^{N_{\mathbf{p}}} \cosh(\pi \sum_{\mathbf{k}}  \mathrm{Re}\left(N_{\mathbf{k}} b_{\mathbf{k}}^{C}\right))    }
\end{align} 
with $b_{\mathbf{k}}^{C} =  \frac{1}{L} \gamma_{\mathbf{k}}^I\left(\gamma_{\mathbf{k}}^R\right)^{-1} \sum_{\mathbf{p} \in C} e^{-i\mathbf{p}\mathbf{k}}$ and

\begin{align}
\expval{\prod_{\mathbf{p} \in C} (-1)^{N_{\mathbf{p}}} \cosh(\pi \sum_{\mathbf{k}} \mathrm{Re}(N_{\mathbf{k}} b_{\mathbf{k}}^{C}))    } &= \frac{\sum_{ \{ N_{\mathbf{k}=\mathbf{0}}=0 \}} \prod_{\mathbf{p} \in C} (-1)^{N_{\mathbf{p}}}  \cosh(\pi \sum_{\mathbf{k}} \mathrm{Re}(N_{\mathbf{k}} b_{\mathbf{k}}^{C}))   e^{2\pi i \sum_{\mathbf{p}} \epsilon_{\mathbf{p}} N_{\mathbf{p}}}   e^{- \pi \sum_{\mathbf{k}} |N_{\mathbf{k}}|^2  \gamma_{\mathbf{k}} }  }{\sum_{ \{ N_{\mathbf{k}=\mathbf{0}}=0 \}} e^{2\pi i  \sum_{\mathbf{p}} \epsilon_{\mathbf{p}} N_{\mathbf{p}}}   e^{- \pi \sum_{\mathbf{k}} |N_{\mathbf{k}}|^2  \gamma_{\mathbf{k}} }}   \nonumber\\
&=\frac{\sum_{ \{ N_{\mathbf{p}}  \}/\sim_1 } e^{-\pi \sum_{\mathbf{k}} \big(|N_\mathbf{k} - \epsilon_{\mathbf{k}} - \frac{1}{2}^{C}_\mathbf{k}|^2 -\frac{1}{4} |b_{\mathbf{k}}^{C}|^2 \big) \gamma_{\mathbf{k}}^{-1} } \cos \bigg( \pi \sum_\mathbf{k} \gamma_{\mathbf{k}}^{-1} \mathrm{Re} \big[ \big(N_\mathbf{k} - \epsilon_\mathbf{k} - \frac{1}{2}^{C}_\mathbf{k}\big) b_{\mathbf{k}}^{C} \big] \bigg)}{\sum_{ \{ N_{\mathbf{p}}  \}/\sim_1 } e^{- \pi \sum_{\mathbf{k}} |N_{\mathbf{k}} - \epsilon_{\mathbf{k}}|^2 \gamma_{\mathbf{k}}^{-1} } }
\end{align}
with $\frac{1}{2}^{C}_\mathbf{k}= \frac{1}{2L} \sum_{\mathbf{p} \in C} e^{-i\mathbf{p}\mathbf{k}}$ and $\gamma_{\mathbf{k}}=\gamma_{\mathbf{k}}^R+ \left(\gamma_{\mathbf{k}}^I\right)^2 \left(\gamma_{\mathbf{k}}^R\right)^{-1}$. Another observable which is used in the manuscript is the electric field. We present for simplicity the expectation value of a horizontal link emanating from vertex $\mathbf{x}$, $\expval{E_{\mathbf{x},1}}$. The plaquette above the link is denoted as $\mathbf{\tilde{p}}$. The expectation value for a vertical link follows analogously.

\begin{align}
\frac{\expval{E_{\mathbf{x},1}}{\Psi_{CPG}}}{\braket{\Psi_{CPG}}}&=\frac{\sum_{ \{ N_{\mathbf{k}=\mathbf{0}}=0 \}}  \sin{(2\pi \sum_{\mathbf{p}} \epsilon_{\mathbf{p}} N_{\mathbf{p}})}  e^{-\pi \sum_{\mathbf{k}} |N_{\mathbf{k}}|^2 \gamma_{\mathbf{k}}} \frac{1}{L} \sum_{\mathbf{k}} \gamma_{\mathbf{k}} \mathrm{Re}\left(N_{\mathbf{k}}  \left( e^{-i\mathbf{k}\mathbf{\tilde{p}}} - e^{-i\mathbf{k}(\mathbf{\tilde{p}}-\mathbf{e_2})}\right) \right)  } {\sum_{ \{ N_{\mathbf{k}=\mathbf{0}}=0 \}}   e^{2\pi i \sum_{\mathbf{p}} \epsilon_{\mathbf{p}} N_{\mathbf{p}}}  e^{-\pi \sum_{\mathbf{k}} |N_{\mathbf{k}}|^2 \gamma_{\mathbf{k}}}} \nonumber\\
&=\epsilon_{\mathbf{\tilde{p}}}-\epsilon_{\mathbf{\tilde{p}}-\mathbf{e_{2}}} + \frac{  \sum_{ \{ N_{\mathbf{p}}  \}/\sim_1 }  e^{- \pi \sum_{\mathbf{k}} |N_{\mathbf{k}} - \epsilon_{\mathbf{k}}|^2 \gamma_{\mathbf{k}}^{-1} } (N_{\mathbf{\tilde{p}-e_{2}}}-N_{\mathbf{\tilde{p}}})  } {\sum_{ \{ N_{\mathbf{p}}  \}/\sim_1 } e^{- \pi \sum_{\mathbf{k}} |N_{\mathbf{k}} - \epsilon_{\mathbf{k}}|^2 \gamma_{\mathbf{k}}^{-1} } } 
\end{align}
The next quantity we present is the gradient of the variational energy with respect to the independent parameters $\gamma_{\mathbf{k}}^R$ and $\gamma_{\mathbf{k}}^I$ ($\mathbf{k} \in \mathcal{K}$). We split the energy into an electric and a magnetic part to make the expressions less cumbersome. We start with the derivatives of the electric energy with respect to $\gamma_{\mathbf{k}}^R$ and $\gamma_{\mathbf{k}}^I$ ($\mathbf{k} \in \mathcal{K}$):

\begin{align}
&\begin{aligned}
&\frac{\partial}{\partial \gamma_{\mathbf{k}}^R} \frac{\expval{H_E}{\Psi_{CPG}}}{\braket{\Psi_{CPG}}}\\
&= \frac{g^2}{4\pi} m_{\mathbf{k}} \left( 1 - \frac{\left(\gamma_{\mathbf{k}}^I\right)^2}{\left(\gamma_{\mathbf{k}}^R\right)^2}\right) \left(4-2 \cos\left(\frac{2\pi k_{1}}{L}\right)-2 \cos\left(\frac{2\pi k_{2}}{L}\right) \right) \\
&\phantom{=}-g^2  m_{\mathbf{k}} \left(\gamma_{\mathbf{k}}^R - \frac{\left(\gamma_{\mathbf{k}}^I\right)^4}{\left(\gamma_{\mathbf{k}}^R\right)^3}  \right)  \left(4-2 \cos\left(\frac{2\pi k_{1}}{L}\right)-2 \cos\left(\frac{2\pi k_{2}}{L}\right) \right) \expval{|N_{\mathbf{k}}|^2}\\
&\phantom{=}+\frac{g^2\pi}{2} m_{\mathbf{k}} \left(1 - \frac{\left(\gamma_{\mathbf{k}}^I\right)^2}{\left(\gamma_{\mathbf{k}}^R\right)^2} \right) \sum_{\mathbf{k'}} \gamma_{\mathbf{k'}}^2  \left(4-2 \cos\left(\frac{2\pi k'_{1}}{L}\right)-2 \cos\left(\frac{2\pi k'_{2}}{L}\right) \right) \left(\expval{|N_{\mathbf{k'}}|^2 |N_{\mathbf{k}}|^2 } -  \expval{|N_{\mathbf{k'}}|^2} \expval{|N_{\mathbf{k}}|^2} \right)
\end{aligned}\\\nonumber\\
&\begin{aligned}
&\frac{\partial}{\partial \gamma_{\mathbf{k}}^I} \frac{\expval{H_E}{\Psi_{CPG}}}{\braket{\Psi_{CPG}}}\\
&= \frac{g^2}{2\pi} m_{\mathbf{k}}  \frac{\gamma_{\mathbf{k}}^I}{\gamma_{\mathbf{k}}^R} \left(4-2 \cos\left(\frac{2\pi k_{1}}{L}\right)-2 \cos\left(\frac{2\pi k_{2}}{L}\right) \right) \\
&\phantom{=}-2g^2  m_{\mathbf{k}} \left(\gamma_{\mathbf{k}}^I + \frac{\left(\gamma_{\mathbf{k}}^I\right)^3}{\left(\gamma_{\mathbf{k}}^R\right)^2}  \right)  \left(4-2 \cos\left(\frac{2\pi k_{1}}{L}\right)-2 \cos\left(\frac{2\pi k_{2}}{L}\right) \right) \expval{|N_{\mathbf{k}}|^2} \\
&\phantom{=}+g^2\pi m_{\mathbf{k}} \frac{\gamma_{\mathbf{k}}^I}{\gamma_{\mathbf{k}}^R}  \sum_{\mathbf{k'}}  \gamma_{\mathbf{k'}}^2  \left(4-2 \cos\left(\frac{2\pi k'_{1}}{L}\right)-2 \cos\left(\frac{2\pi k'_{2}}{L}\right) \right) \left(\expval{|N_{\mathbf{k'}}|^2 |N_{\mathbf{k}}|^2 } -  \expval{|N_{\mathbf{k'}}|^2} \expval{|N_{\mathbf{k}}|^2} \right) \\ 
\end{aligned}
\end{align}  
with 
\begin{align}
\expval{|N_{\mathbf{k'}}|^2 |N_{\mathbf{k}}|^2 } &= \frac{\sum_{ \{ N_{\mathbf{k}=\mathbf{0}}=0 \}} e^{2\pi i \sum_{\mathbf{p}} \epsilon_{\mathbf{p}} N_{\mathbf{p}}} e^{-\pi \sum_{\mathbf{k}} |N_{\mathbf{k}}|^2 \gamma_{\mathbf{k}}}  |N_{\mathbf{k}}|^2  |N_{\mathbf{k'}}|^2 } {\sum_{ \{ N_{\mathbf{k}=\mathbf{0}}=0 \}}   e^{2\pi i \sum_{\mathbf{p}} \epsilon_{\mathbf{p}} N_{\mathbf{p}}}  e^{-\pi \sum_{\mathbf{k}} |N_{\mathbf{k}}|^2 \gamma_{\mathbf{k}}}} \nonumber\\
&= \frac{1}{4\pi^2} \gamma_{\mathbf{k}}^{-1} \gamma_{\mathbf{k'}}^{-1}  +  \gamma_{\mathbf{k}}^{-2} \gamma_{\mathbf{k'}}^{-2} \frac{  \sum_{ \{ N_{\mathbf{p}}  \}/\sim_1 }  e^{- \pi \sum_{\mathbf{k}} |N_{\mathbf{k}} - \epsilon_{\mathbf{k}}|^2 \gamma_{\mathbf{k}}^{-1} } |N_{\mathbf{k}}-\epsilon_{\mathbf{k}}|^2 |N_{\mathbf{k'}}-\epsilon_{\mathbf{k'}}|^2  } {\sum_{ \{ N_{\mathbf{p}}  \}/\sim_1 } e^{- \pi \sum_{\mathbf{k}} |N_{\mathbf{k}} - \epsilon_{\mathbf{k}}|^2 \gamma_{\mathbf{k}}^{-1} } } \nonumber\\
&\phantom{=}- \frac{1}{2\pi} \gamma_{\mathbf{k'}}^{-2} \gamma_{\mathbf{k}}^{-1} \frac{  \sum_{ \{ N_{\mathbf{p}}  \}/\sim_1 }  e^{- \pi \sum_{\mathbf{k}} |N_{\mathbf{k}} - \epsilon_{\mathbf{k}}|^2 \gamma_{\mathbf{k}}^{-1} } |N_{\mathbf{k'}}-\epsilon_{\mathbf{k'}}|^2  } {\sum_{ \{ N_{\mathbf{p}}  \}/\sim_1 } e^{- \pi \sum_{\mathbf{k}} |N_{\mathbf{k}} - \epsilon_{\mathbf{k}}|^2 \gamma_{\mathbf{k}}^{-1} } } \nonumber\\
&\phantom{=}-\frac{1}{2\pi} \gamma_{\mathbf{k}}^{-2} \gamma_{\mathbf{k'}}^{-1} \frac{  \sum_{ \{ N_{\mathbf{p}}  \}/\sim_1 }  e^{- \pi \sum_{\mathbf{k}} |N_{\mathbf{k}} - \epsilon_{\mathbf{k}}|^2 \gamma_{\mathbf{k}}^{-1} } |N_{\mathbf{k}}-\epsilon_{\mathbf{k}}|^2  } {\sum_{ \{ N_{\mathbf{p}}  \}/\sim_1 } e^{- \pi \sum_{\mathbf{k}} |N_{\mathbf{k}} - \epsilon_{\mathbf{k}}|^2 \gamma_{\mathbf{k}}^{-1} } } \nonumber\\
&\phantom{=}+\delta_{\mathbf{k},\mathbf{k'}} \frac{1}{m_{\mathbf{k}}} \left( \frac{1}{2\pi^2}  \gamma_{\mathbf{k}}^{-2} - \frac{2}{\pi} \gamma_{\mathbf{k}}^{-3}  \frac{  \sum_{ \{ N_{\mathbf{p}}  \}/\sim_1 }  e^{- \pi \sum_{\mathbf{k}} |N_{\mathbf{k}} - \epsilon_{\mathbf{k}}|^2 \gamma_{\mathbf{k}}^{-1} } |N_{\mathbf{k}}-\epsilon_{\mathbf{k}}|^2  } {\sum_{ \{ N_{\mathbf{p}}  \}/\sim_1 } e^{- \pi \sum_{\mathbf{k}} |N_{\mathbf{k}} - \epsilon_{\mathbf{k}}|^2 \gamma_{\mathbf{k}}^{-1} } } \right)
\end{align}
We denote by $m_{\mathbf{k}}$ the number of elements in the equivalence class $\mathbf{k} \in K$ which is two if $\mathbf{k} \neq -\mathbf{k}$ and one if $\mathbf{k}=-\mathbf{k}$. The expression for $\expval{|N_{\mathbf{k}}|^2}$ for both high and low $\gamma_{\mathbf{k}}$ approximation can be found in eq.~\eqref{elinfinitesum}. The gradient of the magnetic energy with respect to $\gamma_{\mathbf{k}}^R$ and $\gamma_{\mathbf{k}}^I$ takes the form:
\begin{align}
&\begin{aligned}
&\frac{\partial}{\partial \gamma_{\mathbf{k}}^R} \frac{\expval{H_B}{\Psi_{CPG}}}{\braket{\Psi_{CPG}}}\\  
&=\frac{\pi}{g^2} m_{\mathbf{k}} e^{-\frac{\pi}{4L^2} \sum_{\mathbf{k \neq 0}} \left(\gamma_{\mathbf{k}}^R\right)^{-1}} \sum_{\mathbf{p}} \left[- \frac{1}{4L^2} \left(\gamma_{\mathbf{k}}^R\right)^{-2}   \expval{(-1)^{N_{\mathbf{p}}}  \cosh \left( \pi \sum_{\mathbf{k'}}  \mathrm{Re}(N_{\mathbf{k'}} b_{\mathbf{k'}}^{\mathbf{p}}) \right) }  \right.\\
&\phantom{=}\left. +\frac{1}{L} \frac{\gamma_{\mathbf{k}}^I} {\left(\gamma_{\mathbf{k}}^R\right)^2} \expval{(-1)^{N_{\mathbf{p}}} \mathrm{Re}(N_{\mathbf{k}} e^{-i\mathbf{k}\mathbf{p}}) \sinh\left( \pi \sum_{\mathbf{k'}} \mathrm{Re}(N_{\mathbf{k'}} b_{\mathbf{k'}}^{\mathbf{p}}) \right)   } \right.\\
&\phantom{=}\left. +\left( 1 - \frac{\left(\gamma_{\mathbf{k}}^I\right)^2}{\left(\gamma_{\mathbf{k}}^R\right)^2}\right)  \left (\expval{(-1)^{N_{\mathbf{p}}}  |N_{\mathbf{k}}|^2 \cosh\left(\pi \sum_{\mathbf{k'}} \mathrm{Re}(N_{\mathbf{k'}} b_{\mathbf{k'}}^{\mathbf{p}}) \right)  }  - \expval{(-1)^{N_{\mathbf{p}}} \cosh\left(\pi \sum_{\mathbf{k'}} \mathrm{Re}(N_{\mathbf{k'}} b_{\mathbf{k'}})  \right)} \expval{|N_{\mathbf{k}}|^2} \right) \right]\\ \\
\end{aligned}\\
&\begin{aligned}
&\frac{\partial}{\partial \gamma_{\mathbf{k}}^I} \frac{\expval{H_B}{\Psi_{CPG}}}{\braket{\Psi_{CPG}}}\\  
&=\frac{\pi}{g^2} m_{\mathbf{k}} e^{-\frac{\pi}{4L^2} \sum_{\mathbf{k \neq 0}} \left(\gamma_{\mathbf{k}}^R\right)^{-1}} \sum_{\mathbf{p}} \left[ -\frac{1}{L} \left(\gamma_{\mathbf{k}}^R\right)^{-1}  \expval{(-1)^{N_{\mathbf{p}}} \mathrm{Re}(N_{\mathbf{k}} e^{-i\mathbf{k}\mathbf{p}}) \sinh \left( \pi \sum_{\mathbf{k'}} \mathrm{Re}(N_{\mathbf{k'}} b_{\mathbf{k'}}^{\mathbf{p}}) \right)  } \right.\\
&\phantom{=}\left.+ 2 \frac{\gamma_{\mathbf{k}}^I}{\gamma_{\mathbf{k}}^R} \left(\expval{(-1)^{N_{\mathbf{p}}}  |N_{\mathbf{k}}|^2 \cosh\left(\pi \sum_{\mathbf{k'}} \mathrm{Re}(N_{\mathbf{k'}} b_{\mathbf{k'}}^{\mathbf{p}}) \right)  }  - \expval{(-1)^{N_{\mathbf{p}}} \cosh\left(\pi \sum_{\mathbf{k'}} \mathrm{Re}(N_{\mathbf{k'}} b_{\mathbf{k'}}^{\mathbf{p}})  \right)} \expval{|N_{\mathbf{k}}|^2} \right) \right] 
\end{aligned}
\end{align}
with the usual definition of $b^{\mathbf{p}}_{\mathbf{k}}$ and $\frac{1}{2}_{\mathbf{k}}^{\mathbf{p}}$ from eq.~\eqref{varenergy} and the infinite sums 
\begin{align}
&\expval{(-1)^{N_{\mathbf{p}}}  |N_{\mathbf{k}}|^2 \cosh\left(\pi \sum_{\mathbf{k'}} \mathrm{Re}(N_{\mathbf{k'}} b_{\mathbf{k'}}^{\mathbf{p}}) \right)  }   \nonumber \\
&= \frac{\sum_{ \{ N_{\mathbf{k}=\mathbf{0}}=0 \}} (-1)^{N_{\mathbf{p}}}  |N_{\mathbf{k}}|^2 \cosh\left(\pi \sum_{\mathbf{k'}} \mathrm{Re}\left(N_{\mathbf{k'}} b_{\mathbf{k'}}^{\mathbf{p}}\right)\right)   e^{2\pi i \sum_{\mathbf{p}} \epsilon_{\mathbf{p}} N_{\mathbf{p}}}   e^{- \pi \sum_{\mathbf{k}} |N_{\mathbf{k}}|^2  \gamma_{\mathbf{k}} }  }{\sum_{ \{ N_{\mathbf{k}=\mathbf{0}}=0 \}} e^{2\pi i  \sum_{\mathbf{p}} \epsilon_{\mathbf{p}} N_{\mathbf{p}}}   e^{- \pi \sum_{\mathbf{k}} |N_{\mathbf{k}}|^2  \gamma_{\mathbf{k}} }}   \nonumber\\
&=- \gamma_{\mathbf{k}}^{-2}  \frac{\sum_{ \{ N_{\mathbf{p}}  \}/\sim_1 } e^{-\pi \sum_{\mathbf{k}} \big(|N_\mathbf{k} - \epsilon_{\mathbf{k}} - \frac{1}{2}^{\mathbf{p}}_\mathbf{k}|^2 -\frac{1}{4} |b_{\mathbf{k}}^{\mathbf{p}}|^2 \big) \gamma_{\mathbf{k}}^{-1} }  \sin \left( \pi \sum_\mathbf{k'} \gamma_{\mathbf{k'}}^{-1} \mathrm{Re} \left[ \big(N_\mathbf{k'} - \epsilon_\mathbf{k'} - \frac{1}{2}^{\mathbf{p}}_\mathbf{k'}\big) b_{\mathbf{k'}}^{\mathbf{p}} \right] \right) \mathrm{Re} \left[ \big(N_\mathbf{k} - \epsilon_\mathbf{k} - \frac{1}{2}^{\mathbf{p}}_\mathbf{k}\big) b_{\mathbf{k}}^{\mathbf{p}} \right]}{\sum_{ \{ N_{\mathbf{p}}  \}/\sim_1 } e^{- \pi \sum_{\mathbf{k}} |N_{\mathbf{k}} - \epsilon_{\mathbf{k}}|^2 \gamma_{\mathbf{k}}^{-1} } } \nonumber\\
&\phantom{=}- \gamma_{\mathbf{k}}^{-2} \frac{\sum_{ \{ N_{\mathbf{p}}  \}/\sim_1 } e^{-\pi \sum_{\mathbf{k}} \big(|N_\mathbf{k} - \epsilon_{\mathbf{k}} - \frac{1}{2}^{\mathbf{p}}_\mathbf{k}|^2 -\frac{1}{4} |b_{\mathbf{k}}^{\mathbf{p}}|^2 \big) \gamma_{\mathbf{k}}^{-1} } \cos \left( \pi \sum_\mathbf{k'} \gamma_{\mathbf{k'}}^{-1} \mathrm{Re} \left[ \big(N_\mathbf{k'} - \epsilon_\mathbf{k'} - \frac{1}{2}^{\mathbf{p}}_\mathbf{k'}\big) b_{\mathbf{k'}}^{\mathbf{p}} \right] \right)  \left(|N_\mathbf{k} - \epsilon_{\mathbf{k}} - \frac{1}{2}^{\mathbf{p}}_\mathbf{k}|^2 -\frac{1}{4} |b_{\mathbf{k}}^{\mathbf{p}}|^2 \right)}{\sum_{ \{ N_{\mathbf{p}}  \}/\sim_1 } e^{- \pi \sum_{\mathbf{k}} |N_{\mathbf{k}} - \epsilon_{\mathbf{k}}|^2 \gamma_{\mathbf{k}}^{-1} } }
\end{align}
\begin{align}
&\expval{(-1)^{N_{\mathbf{p}}} \mathrm{Re}(N_{\mathbf{k}} e^{-i\mathbf{k}\mathbf{p}}) \sinh \left( \pi \sum_{\mathbf{k'}} \mathrm{Re}(N_{\mathbf{k'}} b_{\mathbf{k'}}^{\mathbf{p}}) \right)  }  \nonumber\\
&= \frac{\sum_{ \{ N_{\mathbf{k}=\mathbf{0}}=0 \}} (-1)^{N_{\mathbf{p}}} \mathrm{Re}(N_{\mathbf{k}} e^{-i\mathbf{k}\mathbf{p}}) \sinh \left( \pi \sum_{\mathbf{k'}} \mathrm{Re}(N_{\mathbf{k'}} b_{\mathbf{k'}}^{\mathbf{p}}) \right)  e^{2\pi i \sum_{\mathbf{p}} \epsilon_{\mathbf{p}} N_{\mathbf{p}}}   e^{- \pi \sum_{\mathbf{k}} |N_{\mathbf{k}}|^2  \gamma_{\mathbf{k}} }  }{\sum_{ \{ N_{\mathbf{k}=\mathbf{0}}=0 \}} e^{2\pi i \sum_{\mathbf{p}} \epsilon_{\mathbf{p}} N_{\mathbf{p}}}   e^{- \pi \sum_{\mathbf{k}} |N_{\mathbf{k}}|^2  \gamma_{\mathbf{k}} }}   \nonumber\\
&= - \gamma_{\mathbf{k}}^{-1} \frac{\sum_{ \{ N_{\mathbf{p}}  \}/\sim_1 } e^{-\pi \sum_{\mathbf{k}} \big(|N_\mathbf{k} - \epsilon_{\mathbf{k}} - \frac{1}{2}^{\mathbf{p}}_\mathbf{k}|^2 -\frac{1}{4} |b_{\mathbf{k}}^{\mathbf{p}}|^2 \big) \gamma_{\mathbf{k}}^{-1} } \sin \left( \pi \sum_\mathbf{k'} \gamma_{\mathbf{k'}}^{-1} \mathrm{Re} \left[ \big(N_\mathbf{k'} - \epsilon_\mathbf{k'} - \frac{1}{2}^{\mathbf{p}}_\mathbf{k'}\big) b_{\mathbf{k'}}^{\mathbf{p}} \right] \right) \mathrm{Re} \left[ \big(N_\mathbf{k} - \epsilon_\mathbf{k} - \frac{1}{2}^{\mathbf{p}}_\mathbf{k}\big) e^{-i\mathbf{k}\mathbf{p}} \right]}{\sum_{ \{ N_{\mathbf{p}}  \}/\sim_1 } e^{- \pi \sum_{\mathbf{k}} |N_{\mathbf{k}} - \epsilon_{\mathbf{k}}|^2 \gamma_{\mathbf{k}}^{-1} } } \nonumber\\
&\phantom{=}+\frac{1}{2L} \frac{\gamma_{\mathbf{k}}^I}{\left(\gamma_{\mathbf{k}}^R\right)^2+\left(\gamma_{\mathbf{k}}^I\right)^2} \expval{(-1)^{N_{\mathbf{p}}}  \cosh\left(\pi \sum_{\mathbf{k'}} \mathrm{Re}(N_{\mathbf{k'}} b_{\mathbf{k'}}^{\mathbf{p}}) \right)  } 
\end{align}
The expression for $\expval{(-1)^{N_{\mathbf{p}}}  \cosh\left(\pi \sum_{\mathbf{k'}} \mathrm{Re}(N_{\mathbf{k'}} b_{\mathbf{k'}}^{\mathbf{p}}) \right)  }$ can be found in eq.~\eqref{maginfinitesum}.
A crucial quantity for the time-dependent variational principle is the Gram matrix. It is defined as the overlap between two tangent vectors on the variational manifold. Therefore, it is not only the overlap between the derivatives of the ansatz with respect to the variational parameters but it also needs to be projected onto the variational manifold (see eq.~\eqref{projectionvarmanifold}):
\begin{equation}
\begin{aligned}
G_{\mathbf{k}\mathbf{k'}}=& \frac{\pi^2}{4} m_{\mathbf{k}} m_{\mathbf{k'}} \left(\expval{|N_{\mathbf{k}}|^2 |N_{\mathbf{k'}}|^2} - \expval{|N_{\mathbf{k}}|^2} \expval{|N_{\mathbf{k'}}|^2} \right) \left[ \left(1 -\frac{\left(\gamma_{\mathbf{k}}^I\right)^2} {\left(\gamma_{\mathbf{k}}^R\right)^2} \right) 
\left(1 -\frac{\left(\gamma_{\mathbf{k'}}^I\right)^2} {\left(\gamma_{\mathbf{k'}}^R\right)^2} \right)-4 \frac{\gamma_{\mathbf{k}}^I}{\gamma_{\mathbf{k}}^R} \frac{\gamma_{\mathbf{k'}}^I}{\gamma_{\mathbf{k'}}^R}+2i  \frac{\gamma_{\mathbf{k}}^I}{\gamma_{\mathbf{k}}^R} \left(1 -\frac{\left(\gamma_{\mathbf{k'}}^I\right)^2} {\left(\gamma_{\mathbf{k'}}^R\right)^2} \right) \right.\\
&\left.-2i \left(1 -\frac{\left(\gamma_{\mathbf{k}}^I\right)^2} {\left(\gamma_{\mathbf{k}}^R\right)^2} \right) \frac{\gamma_{\mathbf{k'}}^I}{\gamma_{\mathbf{k'}}^R} \right] + \delta_{\mathbf{k},\mathbf{k'}} m_{\mathbf{k}} \left( \frac{1}{8 \left(\gamma_{\mathbf{k}}^R\right)^2} -\frac{\pi}{2} \expval{|N_{\mathbf{k}}|^2} \left( \frac{1 }{\gamma_{\mathbf{k}}^R} + \frac{\left(\gamma_{\mathbf{k}}^I\right)^2}{\left(\gamma_{\mathbf{k}}^R\right)^3} \right)\right)\\
\end{aligned}
\end{equation}
\end{widetext}
\end{document}